\documentclass[acmsmall,10pt]{acmart}

\usepackage[normalem]{ulem}

\usepackage{booktabs}   %
\usepackage{subcaption} %
\usepackage{xspace}
\usepackage{listings}
\usepackage{cleveref}

\usepackage{color}

\usepackage[title]{appendix}

\usepackage{algorithm}
\usepackage{fixltx2e}
\usepackage{algpseudocode}
\usepackage{makecell}

\usepackage[english]{babel}
\usepackage[utf8x]{inputenc}
\usepackage{anyfontsize}
\usepackage{wrapfig}

\usepackage{amsmath}
\DeclareMathOperator*{\argmax}{argmax}
\DeclareMathOperator*{\argmin}{argmin}

\MakeRobust{\Call}
\algrenewcommand\algorithmicrequire{\textbf{Input:}}
\algrenewcommand\algorithmicensure{\textbf{Output:}}

\newif\ifcommentary
\commentarytrue

\newif\ifanon
\anonfalse

\ifcommentary
    
    \newcommand{\thurston}[1]{\textbf{\color{blue}{Thurston: #1}}}
    \newcommand{\jose}[1]{\textbf{\color{green}{Jose: #1}}}
    \newcommand{\nv}[1]{\textbf{\color{darkblue}{(Nikos: #1)}}}
\else
    \newcommand{\thurston}[1]{}
    \newcommand{\jose}[1]{}
    \newcommand{\nv}[1]{}
\fi

\def\eg{{\em e.g.},\xspace}
\def\ie{{\em i.e.},\xspace}
\def\etc{{\em etc.}\xspace}
\def\vs{{\em vs.}\xspace}

\newcommand{\slime}{\textsc{BIEBER}\xspace}

\newcommand{\stbimage}{\texttt{stb\_image}\xspace}
\newcommand{\catimg}{\texttt{catimg}\xspace}
\newcommand{\diode}{\texttt{DIODE}\xspace}
\newcommand{\sndlib}{\texttt{SndLib}\xspace}
\newcommand{\smtlib}{\texttt{SMT-LIB}\xspace}
\newcommand{\openwrt}{\texttt{OpenWrt}\xspace}

\newcommand{\naive}{na\"ive\xspace}
\newcommand{\code}[1]{\texttt{#1}}

\newcommand{\pluseq}{\mathrel{+}=}
\newcommand{\minuseq}{\mathrel{-}=}
\newcommand{\blobformat}{\ttfamily\small}

\newcommand{\elcheaposubsection}[1]{\subsection{#1}}

\usepackage{hyperref}

\setcopyright{none}

\acmConference[ARXIV '21]{ARXIV 21'}{April 20, 2021}{Cambridge, MA}
\acmBooktitle{ARXIV '21: April 20, 2021, Cambridge, MA}
\acmPrice{3.00}
\acmISBN{999-9-9999-XXXX-X/99/99}

\begin{document}

\title[Inferring Drop-in Binary Parsers from Program Executions]{Inferring Drop-in Binary Parsers from Program Executions}

\author{Thurston H. Y. Dang}
\affiliation{%
  \institution{Massachusetts Institute of Technology}
  \country{USA}}
  
\author{Jos\'{e} P. Cambronero}
\affiliation{%
  \institution{Massachusetts Institute of Technology}
  \country{USA}}

\author{Martin C. Rinard}
\affiliation{%
  \institution{Massachusetts Institute of Technology}
  \country{USA}}

\date{}

\begin{abstract}

We present \slime (Byte-IdEntical Binary parsER), the first system to model and regenerate a full working parser from instrumented program executions.
To achieve this, \slime exploits the regularity (\eg header fields and array-like data structures) that is commonly found in file formats. Key generalization steps derive strided loops that parse input file data and rewrite concrete loop bounds with expressions over input file header bytes. These steps enable \slime to generalize parses of specific input files to obtain parsers that operate over input files of arbitrary size.  \slime also incrementally and efficiently infers a decision tree that reads file header bytes to route input files of different types to inferred parsers of the appropriate type. The inferred parsers and decision tree are expressed in an intermediate language that is independent of the original program; separate backends (C and Perl in our prototype) can translate the intermediate representation into the same language as the original program (for a safer drop-in replacement), or automatically port to a different language.
An empirical evaluation shows that \slime can successfully regenerate parsers for six file formats (waveform audio [1654 files],  MT76x0 .BIN firmware containers [5 files], OS/2 1.x bitmap images [9 files], Windows 3.x bitmaps [9971 files], Windows 95/NT4 bitmaps [133 files], and Windows 98/2000 bitmaps [859 files]), correctly parsing 100\% ($\geq$ 99.98\% when using standard held-out cross-validation) of the corresponding corpora. %
The regenerated parsers contain automatically inserted safety checks that eliminate common classes of errors such as memory errors. We find that \slime can help reverse-engineer file formats, because it automatically identifies predicates for the decision tree that
relate to key semantics of the file format. We also discuss how \slime helped us detect and fix two new bugs in \stbimage as well as independently rediscover and fix a known bug.

\end{abstract}

\maketitle

\thispagestyle{empty}

\label{sec:introduction}
\section{Introduction}

Input file parsers can be difficult to fully debug---they are typically expected to process arbitrary inputs, potentially from malicious sources, and as a result can miss rare corner cases. These missed cases can cause incorrect parses or even make the parser vulnerable to attacks~\citep{InputRectification, RecoveryShepherding, CodePhage, FilteredIterator}.
Motivated by the observation that input file parsers usually process typical inputs correctly,  we have developed a new technique that first
processes data from instrumented executions of the program to build a model of the behavior of the parser (specifically, the translation from input file bytes to the data structures that hold the parsed data), and then uses the model to {\em regenerate} a new version of the parser, in a target language that may be different from the original program.

We have implemented this technique in the \slime system. \slime ingests log files that map expressions over input file bytes to the values stored in the data structures that hold the parsed data; these log files may, for example, be generated by running an off-the-shelf instrumented version of an existing parser over a corpus of input files. Each log file records the specific mapping from input file bytes to data structure contents that the existing parser produced for the input at hand. This mapping contains \emph{no} information about the internal control flow or looping structure that the parser used to read the file, and \slime does not exploit or require any such information. \slime can therefore work with any existing
parsers for our in-scope file formats (Section \ref{sec:in_scope}), provided they correctly parse the corpus of input files. Since each mapping is specific to the type and size of the file run through the instrumented parser, \slime processes log files from executions on multiple input files, and then integrates information from these multiple examples to obtain a regenerated parser decision tree that works on files of varying types and sizes. \slime implements several key techniques:
\begin{itemize}
    \item {\bf Loop Summarization:} \slime's input is a flat listing of the data structure values as derived from the input file bytes, which is only accurate for files of the exact type and size as the file on which the instrumented application was run. To facilitate loop generalization, \slime first infers a loop (or set(s) of nested loops) that implement the specific translation for each input file. Key aspects of the algorithm include the ability to work with parsers that handle strided input or reorder input file bytes as they move into the data structures.

    \item {\bf Loop Generalization:} \slime's input does not contain information about loops, let alone how the loop bounds are derived. To produce parsers that generalize to input files of different sizes, \slime correlates values in the concrete loop bounds with values in the input file header, to infer bounds expressions that generalize these loops to files of different sizes.

    \item {\bf Parsing Multiple File Types via a Decision Tree:} Input files often have different types, requiring distinct parsing strategies; for example, waveform audio may store samples with different bit-depths or number of channels. To create a combined parser that can works for all of the input file types, \slime generates a decision tree that incorporates inferred predicates on relevant input file bytes to route each file to a parser of the corresponding type.

    \item {\bf Iterative Generalization:} To drive down the overhead associated with executing instrumented versions of the existing parser, \slime exploits file size and execution time properties. First, small input files require much less time to process than large input files. \slime therefore processes the files in the test corpus in order from smallest to largest.
    Second, \slime's regenerated parser takes much less time to execute than the instrumented version of the existing parser. \slime therefore adopts an optimized technique that first executes the regenerated parser (without instrumentation) on the next file in turn. Only if the parse fails does \slime then execute the instrumented version (with associated overhead) of the existing parser.

    Our results show that our optimizations significantly reduce the number of instrumented executions. For example, to build the complete decision tree parser (consisting of 20 leaves/type-specific parsers) for the corpus of 11,008 bitmap files, \slime only executes the instrumented version of the application for 74 input files.

    \item {\bf Improved Code through Regeneration:} \slime's algorithms output an intermediate representation that can be translated to different languages by using an appropriate backend; the regenerated code need not be the same language as the original program. This allows convenient ``porting'' of the parser to different languages (C and Perl in \slime's prototype).
    Moreover, if the target language is not inherently memory-safe, \slime enforces memory safety by passing all writes to the arrays that hold the parsed data through a helper function.\footnote{Use-after-free is prevented by construction in the regeneration process.} This helper function performs all required bounds checks; resizable arrays support inputs of arbitrary sizes.
\end{itemize}

\slime also helped identify inputs that triggered corner case bugs: one of our test applications, a bitmap parser, contained two previously unknown bugs that caused some (uncommon case) bitmaps to parse (and therefore display) incorrectly.\footnote{We have reported these to the developers, and one has since been fixed.} The \slime-generated decision tree contained a distinct parser (or parsers, if the parsers for the buggy files failed to generalize) for each of the classes of inputs that trigger each bug --- it routed all input files that trigger the bug, and only those files, to the corresponding parser. This isolation of the bug-triggering inputs helped us discover and correct these bugs. When we  reran \slime on the corrected application, \slime generated a simplified decision tree that grouped these inputs together with other inputs that did not trigger the bugs. The decision tree automatically identified predicates that are meaningful to the file format semantics -- without using the file specification -- which we found useful for surfacing anomalies in the existing parser and enhancing our understanding of the file format.

\elcheaposubsection{Contributions}
\slime is the first system to model and regenerate a full working parser from instrumented program executions.
Because the regenerated parser produces data structures that are byte-for-byte identical to the data structures in the original program, the regenerated parser comprises an immediate drop-in replacement for the original parser. 
\slime does \emph{not} use control-flow information, and instead leverages a combination of  inference techniques to automatically extract structured properties of the file format.

This paper presents:
\begin{itemize}
   \item {\bf Algorithms for Modeling and Regenerating Parsers:} We present the aforementioned techniques of loop summarization, loop generalization, parsing multiple file types via a decision tree, iterative generalization, and improved code through regeneration.
   \item {\bf Empirical Evaluation:} We have implemented those algorithms in the \slime system. \slime generates working code --- whereby the target language can be chosen independently of the language of the original code, simply by selecting a different backend --- that includes systematically generated checks where necessary. We evaluate \slime on six file formats: waveform audio,  MT76x0 .BIN firmware containers, OS/2 1.x bitmaps, Windows 3.x bitmaps, Windows 95/NT4 bitmaps, and Windows 98/2000 bitmaps. The results show that \slime efficiently models and regenerates parsers that successfully parse all input file types represented in the training set, generalizing to many files beyond those in the training set. Additionally, the decision trees built by \slime have automatically identified predicates that are important to the file formats. We also discuss how \slime assisted in uncovering two previously unreported bugs in \stbimage.
\end{itemize}

\begin{figure}[!h]
\centering
\includegraphics[scale=0.90]{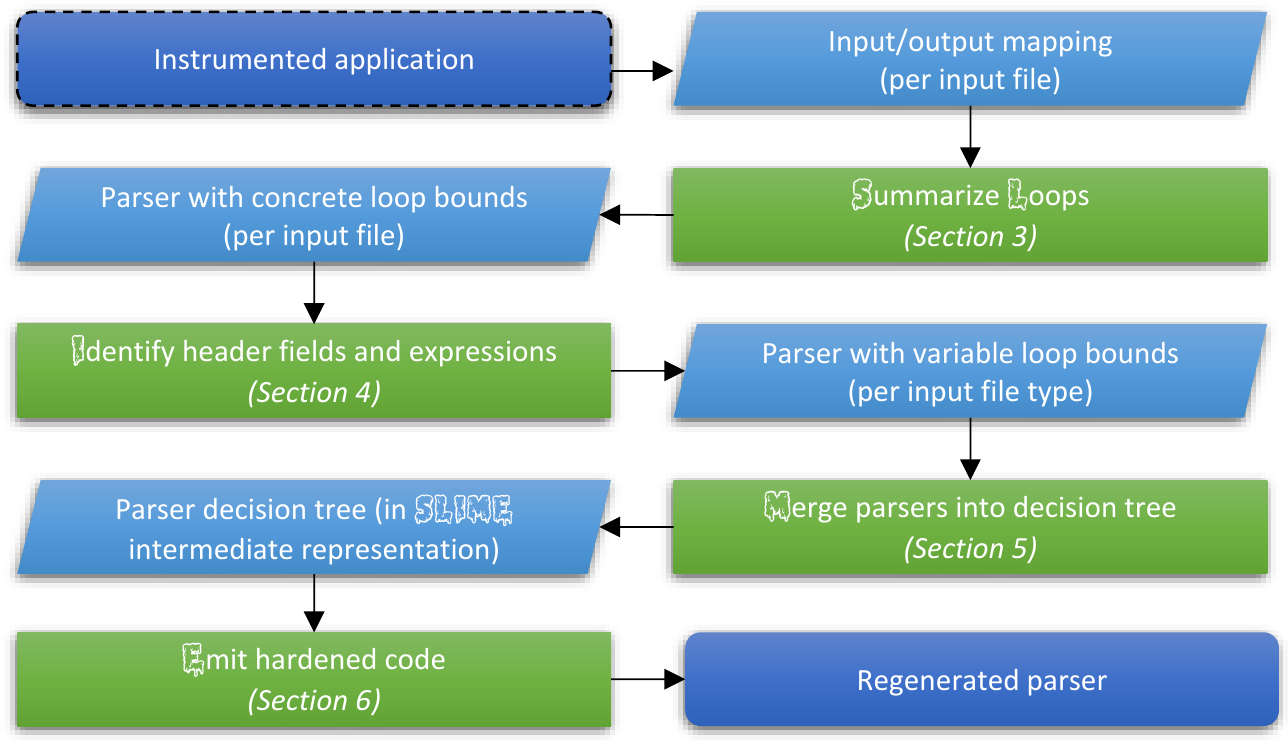}
\caption{
The \slime pipeline.
}
\label{fig:slime-pipeline}
\end{figure}

\elcheaposubsection{Structure of the Paper}
Section~\ref{sec:instrumentation} describes the binary data formats that are in scope for \slime, along with the expected format of the instrumentation logs.
The subsequent four sections describe \slime's key algorithms (Figure \ref{fig:slime-pipeline}):
Section~\ref{sec:summarization} presents loop summarization algorithms, which infers structured for-loops from the flat instrumentation logs;
Section~\ref{sec:identify_header} presents
our algorithm to identify and rewrite key constants to more general
expressions, generalizing our parsers along key dimensions;
Section~\ref{sec:parser} introduces a recursive algorithm
to build a parser decision tree that can correctly parse inputs
of different types, with optimizations to reduce training time; and
Section~\ref{sec:backend} discusses \slime's C and Perl backends.
Section~\ref{sec:evaluation} describes the methodology used to evaluate \slime's ability to regenerate parsers, improve security, provide insight into how the parser processes a file format, and discover bugs; Section~\ref{sec:results} presents the results.
Section~\ref{sec:related_work} describes related work, and Section~\ref{sec:conclusion} concludes.

\section{File Formats}
\label{sec:instrumentation}

\elcheaposubsection{In-scope File Formats}
\label{sec:in_scope}

\slime infers binary data formats that may contain several features, individually or in combination.
We now outline the challenges associated with each of these and the advantages
of using \slime to tackle them.

\noindent\textbf{Starting point:} We assume we have only one ``type''\footnote{``type'' will be precisely defined in Section \ref{sec:parse-by-type}. For now, our intuitive understanding will suffice: for example, monochrome vs. 24-bit color images are different file types, but the width/height of the image does not affect the type.}, with fixed-length input and output buffer, where each output byte is a deterministic function of one or more input bytes. For example, early versions of the ICO icon format (monochrome images with a fixed width/height of 16x16 pixels) meet this definition.
For these cases, header bytes are not necessary for the parser to identify the
``type'' of the file (since we assume there is only one type) nor the length of
any data (since the input and output are fixed-length). Nonetheless, file formats with these features suffice to demonstrate many of \slime's capabilities and benefits: regenerating a compilable parser from execution
logs (encoded using Z3 expressions: see Appendix \ref{sec:diode_log}) requires significant processing by \slime; moreover, regeneration provides benefits, such as porting to a different language.

\noindent\textbf{Enhancement A: Variable length ``chunks''. } File formats can be enriched by allowing variable-length data, with the length stored in the header. This is similar to the Resource Interchange File Format (RIFF)~\cite{riff} ``family'' of file formats. Parsers for these formats are often vulnerable to a common class of bugs (buffer overflows), since the size of the output buffer may not be correctly computed. For example, CVE-2013-2028 is ``chunked Transfer-Encoding request with a large chunk size''~\cite{chunked_cve}.

The instrumentation logs used by \slime do not contain information about the
original looping structure in the parser. This looping
structure is needed to generalize over varying sizes.
\slime therefore introduces loop summarization (Section~\ref{sec:summarization})
and expression rewriting (Section \ref{sec:identify_header}) to infer this structure.
Without those algorithms, a parser would conflate varying file sizes
with varying file types, thus failing to generalize to unseen sizes and resulting in
special-casing each unique size observed.

\noindent\textbf{Enhancement B: Multiple file types.} File formats can support multiple types (for example,  mono \vs stereo audio files), which have to be distinguished based on expressions over the header bytes.
This complication is orthogonal to Enhancement A, as varying file formats could be present
with fixed lengths.  Notice that, even though in the simplest case each file type
has a fixed length, buffer overflows are still possible, as the parser
can misidentify the file type (resulting in an incorrect fixed length).
Similarly to Enhancement A, given that the \slime instrumentation log files do not include
control-flow information, \slime has to infer the relevant header bytes
that indicate the file type. To tackle this challenge, \slime
infers file-type separating predicates as part of its decision tree
algorithm (Section~\ref{sec:parser}).

\noindent\textbf{Composition of Enhancements.} Enhancements A and B are orthogonal: we can have multiple fixed-length types, or a single variable-length type. \slime can also handle file formats that contain both enhancements \ie multiple variable-length types. This is significantly harder than solving either enhancement in isolation. If we only had Enhancement A (only one type of file, but with variable-length chunks), there is less ambiguity about the header bytes since, by definition of a single file type, all the inputs share the same header format.
\slime's algorithms (Section \ref{sec:identify_header}) handle the more complicated case of identifying header fields when there are multiple file types.

\elcheaposubsection{Out-of-scope File Formats}
The features that we have selected are a \emph{subset} of the features found in
many common file formats, and are sufficient for \slime to
regenerate parsers for several commonly used file formats (see Section
\ref{sec:evaluation}). However, these features are not sufficient to parse
\emph{all} file formats that a user may be interested in.
Nonetheless, we believe that the principles presented here are illustrative
of the approaches required to extend \slime with additional features, which would allow
regenerating parsers for a broader range of formats.

\elcheaposubsection{Instrumentation Log Format}

The input to \slime{}'s pipeline is a mapping from input file bytes to output data structures that hold the parsed data. Our \slime prototype uses \diode~\cite{diode}, an off-the-shelf instrumentation tool for byte-level data-flow tracking.
\slime does not depend on how the mapping is obtained, but as shorthand, we will refer to the mapping files as ``\diode logs''.

Given a concrete input file and an instrumented parser, \diode emits
symbolic expressions (in Z3 format~\cite{z3}) for the output variables as a function of the input bytes.
These symbolic expressions capture only the mapping generated by that particular execution and thus only describe the execution for inputs of the exact same
type and size as the file parsed by the instrumented parser.
The logs are, after some pre-processing, of the form
$\text{out}[i] = f(\text{in}[j_1],\ldots,\text{in}[j_n])$, where \textit{out} corresponds to a data structure,
\textit{in} corresponds to the input file pointer,
$i$, $j_1$, \ldots, $j_n$ are offsets such that $x[i]$ refers to the byte
at position $i$ in $x$, and $f(\cdot)$ is some byte-level expression over
its arguments.
See Appendix \ref{sec:diode_log} for an example of a \diode log file and Appendix \ref{sec:example_diode} for an example of a Z3 symbolic expression.

\section{Summarize Loops}
\label{sec:summarization}

The input-output index mappings produced by \diode
describe the translation between input-output bytes for a single input type (for example, an audio file with a fixed number of channels and samples, or an image of fixed width, height, and bit-depth). We therefore need to generalize the mapping to work with inputs of variable length.

This step rewrites the input-output
index mappings into a succinct form where \slime can, at a later stage, alter a key dimension:
size. By summarizing the input-output index mappings into a loop, we can subsequently induce
parsing of new inputs by rewriting the loop bounds as an expression of header field bytes (Section \ref{sec:identify_header}).

\noindent\textbf{Basic summarization algorithm.} \slime performs linear interpolation using the first two
output-input index pairs. For example, with index pairs of: \begin{lstlisting}[basicstyle=\blobformat, escapeinside={<@}{@>}]
    out [<@\textbf{0}@>] = in [<@\textbf{44}@>]; out [<@\textbf{1}@>] = in [<@\textbf{46}@>]; out [2] = in [48];
    ... out [9] = in [60];
\end{lstlisting}
\slime interpolates the first two ``points'', $(0,44)$ and $(1,46)$, to obtain:
\begin{lstlisting}[basicstyle=\blobformat, escapeinside={<@}{@>}]
    inputIndex = (outputIndex * 2) + 44
\end{lstlisting}

\slime then generates a \code{for} loop, with the upper-bound set to match as many output-input index pairs as possible (in this example, \code{outputIndex} $\in [0 .. 9]$). For bottom-up bitmaps, one iteration of this process will capture only a single row of pixels; thus,
\slime repeats this interpolation until all index pairs are part of a loop body (for example, one \code{for} loop per row of pixels, in the case of bottom-up bitmaps). \slime then applies the same process to the \texttt{for} loops' bounds to create nested loops (\eg a 2D loop for bottom-up bitmaps),
until a fixed point is reached. Linear interpolation always matches at least the two points that the line was fit through, thus guaranteeing
convergence.

\noindent\textbf{Finding the optimal stride.} The ``stride'' parameter controls how many bytes of input are processed in each iteration of an innermost \code{for} loop. Conceptually, \slime treats the input \code{in[i]} as different widths according to the stride; for example, if the stride is 3 bytes, then \code{in[2]} consists of bytes \code{6..8} of the input.

The basic summarization algorithm generates loops that accurately describe the
concrete input/output index mapping, regardless of the chosen stride value. To determine
an appropriate stride, we rely on a parsimony heuristic to choose the stride
that results in the fewest bytes of generated intermediate representation.
Our intuition is that the
most compact representation is more likely to group together
logically related output bytes.

\elcheaposubsection{Case Study 1: Waveform Audio Parser}
\label{sec:summarize_wav}
Figure \ref{fig:parser_WAV} shows an excerpt of the parser (left channel, least-significant byte) generated for a 16-bit stereo audio file.
Note that \slime's algorithms generate an intermediate representation that can be converted to different languages by selecting the appropriate backend (C or Perl in our prototype).

\begin{figure}[h!]
\begin{lstlisting}[basicstyle=\blobformat, escapeinside={<@}{@>}]
MIN_X := 7;
MIN_Y := 44;
MIN_Y0_0 := MIN_Y + 0;
LOOP_BOUND_A := 19840; <@\textcolor{purple}{// readWord\_32le\_s (fp, 40);}@>
FACTOR_B_0 := 2;
for (idxA := 0; idxA < LOOP_BOUND_A; idxA += 4) {
    NUMERIC_A_0 := indexA * FACTOR_B_0;
    NUMERIC_A_1 := indexA;
    x0 := NUMERIC_A_0 + MIN_X;
    y0_0 := NUMERIC_A_1 + MIN_Y0_0;
    DATA_STRUCTURE [x0] := DIODE_EXPR (y0_0, file); }
\end{lstlisting}
\caption{
Excerpt from a \slime generated parser for a 16-bit stereo audio file. The rewritten constant, shown in the comments,
would generalize the parser to inputs of different length.
We abstract out the repeated expression over input
bytes as \lstinline{DIODE_EXPR}.
}
\label{fig:parser_WAV}
\end{figure}

\elcheaposubsection{Case Study 2: Bitmap Parser}
\label{sec:summarize_bmp}
Figure \ref{fig:parser_61x76_BGR} shows the parser that is generated for a 61x76
24-bit (BGR) bottom-up bitmap. The summarization algorithm identifies how the input format (bottom-up, BGR, rows padded to a multiple
of four bytes) is transformed to \stbimage's output format (top-down, RGB, no padding).

\begin{figure}[!h]
\begin{lstlisting}[basicstyle=\blobformat, escapeinside={<@}{@>}]
MIN_Y := 54; // <@\textcolor{purple}{readWord\_32le\_s (fp, 10);}@>
MIN_Y0_0 := MIN_Y + 2;
MIN_X := 0;
LOOP_BOUND_A := 61; <@\textcolor{purple}{// readWord\_32le\_s (fp, 18);}@>
LOOP_BOUND_B := 76; <@\textcolor{purple}{// readWord\_32le\_s (fp, 22);}@>
FACTOR_B_0 := 3;
FACTOR_B_2 := 3;
FACTOR_C_0 := 183; <@\textcolor{cyan}{// LOOP\_BOUND\_A * FACTOR\_B\_0;}@>
FACTOR_C_1 := -184; <@\textcolor{cyan}{//   pad4(-LOOP\_BOUND\_A * FACTOR\_B\_0);}@>
ADDEND_C_1 := -75; <@\textcolor{cyan}{// (-LOOP\_BOUND\_B + 1);}@>
for (idxB := 0; idxB < LOOP_BOUND_B; idxB++) {
  NUM_B_1 := idxB * FACTOR_C_0;
  NUM_B_3 := (idxB + ADDEND_C_1) * FACTOR_C_1;
  for (idxA := 0; idxA < LOOP_BOUND_A; idxA++) {
    NUM_A_0 := idxA * FACTOR_B_0 + NUM_B_1;
    NUM_A_1 := idxA * FACTOR_B_2 + NUM_B_3;
    x0 := NUM_A_0 + MIN_X; y0_0 := NUM_A_1 + MIN_Y0_0;
    DATA_STRUCTURE [x0] := DIODE_EXPR (y0_0, file)); // <@\textcolor{blue}{Blue}@> subpixel
    x1 := x0 + 1; y1_0 := y0_0 + (-1);
    DATA_STRUCTURE [x1] := DIODE_EXPR (y1_0, file)); // <@\textcolor{green}{Green}@> subpixel
    x2 := x0 + 2; y2_0 := y0_0 + (-2);
    DATA_STRUCTURE [x2] := DIODE_EXPR (y2_0, file)); // <@\textcolor{red}{Red}@> subpixel } }
\end{lstlisting}
\caption{
\slime's generated parser for a 61x76 BGR bottom-up bitmap. The rewritten constants, shown in the comments,
generalize the parser to inputs of different width, height and bitmap version.
\slime abstracts out a repeated expression over input
file bytes as \lstinline{DIODE_EXPR}.
}
\label{fig:parser_61x76_BGR}
\end{figure}

\begin{figure}
\centering
\includegraphics[width=0.75\textwidth]{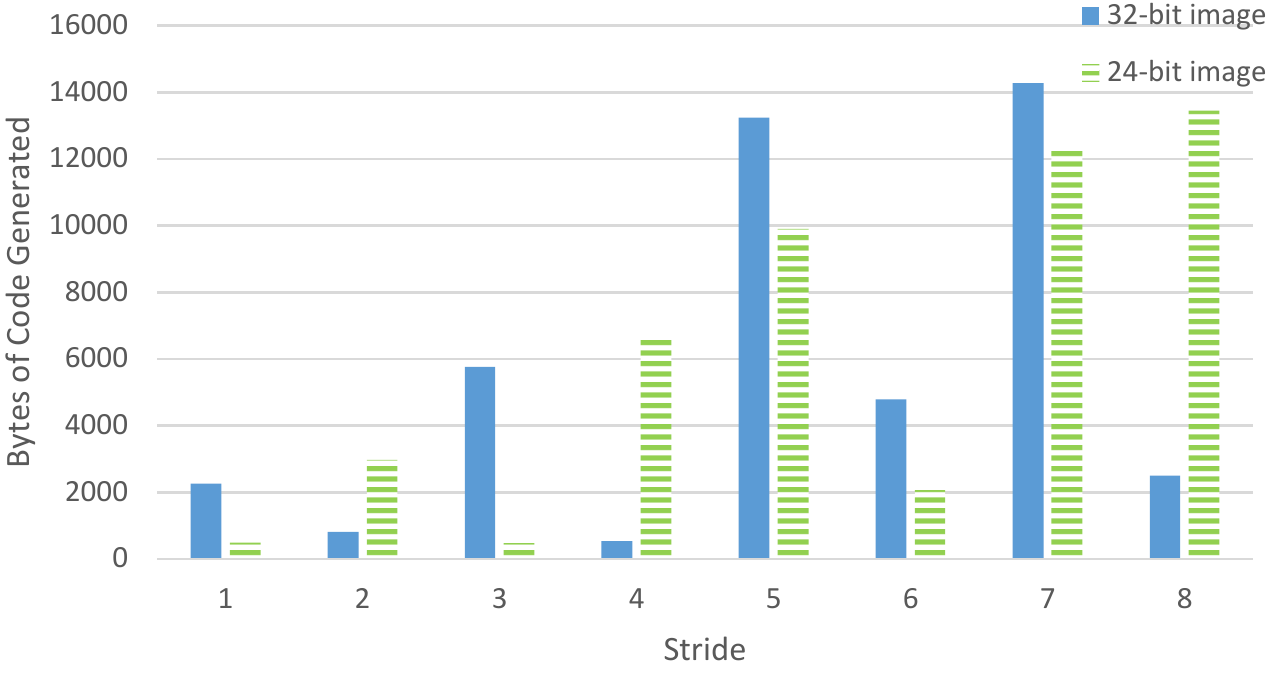}
\caption{
\textbf{Stride detection.}
Number of bytes of code generated for each stride, for 23x73 24- and 32-bit bottom-up bitmaps.
Our parsimony heuristic can correctly identify the optimal strides of 3 and 4 bytes respectively.
}
\label{fig:number-of-bytes-per-stride-23x73}
\end{figure}

The stride of 3 bytes for the BGR parser is encoded as \code{FACTOR\_B\_2} in Figure \ref{fig:parser_61x76_BGR}. Figure \ref{fig:number-of-bytes-per-stride-23x73} shows that the optimal stride of 3 for BGR images was automatically detected by our parsimony heuristic.
A stride of 1 byte also results in reasonably compact code for BGR images; this is because, to reverse
the RGB $\Leftrightarrow$ BGR subpixels, \slime can choose between:

\lstinline[basicstyle=\blobformat, escapeinside={<@}{@>}]!out[i] = in[i+2];    out[i+1] = in[i+1];    out[i+2] = in[i];!

\noindent or:
\lstinline[basicstyle=\blobformat, escapeinside={<@}{@>}]!for (j = 0; j < 3; j++) { out [i+j] = in [i+(2-j)]; }!

For 32-bit bottom-up images, \slime correctly identifies that 4 bytes is the best stride (Figure \ref{fig:number-of-bytes-per-stride-23x73}).

\section{Identify header fields and expressions}
\label{sec:identify_header}

Loop summarization created a set of \texttt{for} loops representing the
output-input index mapping for each expression, with fixed loop bounds.
A parser derived from this mapping will work only for inputs of the same size (\eg audio length or image width/height) and type as the training example.

To generalize the parser to work across inputs of different size (but same type), we
rewrite a subset of
constant definitions as arithmetic expressions over key constants, and then replace key constants with reads over header fields. This rewriting process is applied to \slime's intermediate representation, thus it is independent of the target language.

\elcheaposubsection{Rewrite constants as expressions}
\label{sec:rewrite_constants}

\slime first identifies the set of constants that are candidates for rewriting in the
generated parser IR (\eg Figures \ref{fig:parser_WAV} and \ref{fig:parser_61x76_BGR}). The code
generated by the loop summarization stage was designed to facilitate this
rewrite: for-loops are simplified to be zero-indexed with use of the
$<$ and a simple constant loop bound (which can represent key format properties --- such as length for audio or width/height for bitmaps --- which can be read from the input file), with
traversal complexity shifted to the \code{FACTOR}, \code{ADDEND} and
\code{MIN\_Y}/\code{MIN\_Y0\_0} variables.
Only \code{FACTOR} and \code{ADDEND}
are candidates for rewriting; \code{MIN\_X}, \code{MIN\_Y} and \code{MIN\_Y0\_0}
are simple offsets into the output and input data. \slime generates candidate expressions by instantiating a fixed set of templates
with variables. These templates were designed to cover common file format properties, but can be easily extended:

\begin{lstlisting}[basicstyle=\blobformat, escapeinside={<@}{@>}]
    -x + 1 // upside-down or reverse
          x * y,        -x * y
    pad4 (x * y), pad4 (-x * y)  // word-aligned

\end{lstlisting}

When rewriting expressions in loops, it is often the case that inner nested
loops depend on constants from outer loops \eg \code{FACTOR\_C\_0}
\code{=} \code{LOOP\_BOUND\_A} \code{*} \code{FACTOR\_B\_0}; for example, with bottom-up bitmaps, the outer loop keeps
track of the height, and code inside the inner loop uses both the width and
height to calculate the offset of each subpixel. \slime therefore
restricts the new expressions to use only variables that are in-scope
(\ie at an outer nesting level).

\noindent\textbf{Resolving ambiguity through voting.}
In Figure \ref{fig:parser_61x76_BGR}, given:
\begin{lstlisting}[basicstyle=\blobformat, escapeinside={<@}{@>}]
LOOP_BOUND_A = 61;  LOOP_BOUND_B = 76;  FACTOR_B_0 = 3;  FACTOR_C_0 = 183;
\end{lstlisting}
\code{FACTOR\_C\_0} is \code{LOOP\_BOUND\_A} \code{*} \code{FACTOR\_B\_0}
is the only \emph{parsimonious} rewrite produced by our templates.
Often, however, constant definitions can be rewritten in many different ways, especially if the input dimensions are small or
indistinguishable (\eg for square bitmaps, the width field \code{LOOP\_BOUND\_A} and height field \code{LOOP\_BOUND\_B} are equal).

To handle competing rewrites, \slime employs a plurality-based voting scheme that integrates information from multiple exemplar files.
For each input file, \slime computes the n-fold Cartesian product of candidate
constant rewrites for each variable, and casts one vote per n-tuple.
\noindent \slime then chooses the n-tuple with the most votes across
all input files. If there are multiple equally popular n-tuples, \slime again opts
for parsimony, and chooses the n-tuple with the
shortest aggregate string representation (\eg if \texttt{pad4(x)} and \texttt{x} are equally
popular, \slime chooses \texttt{x}). Any further ties are broken arbitrarily.
The parser created using this n-tuple will, by construction, work for
at least one input file.
The n-fold Cartesian product used in this voting scheme has, in the worst-case,
exponential time and memory complexity. However, it is tractable in practice because of the small number of variables involved, in part because \slime's loop summarization step minimizes the number of loops. We discuss polynomial-time algorithms in Appendix \ref{sec:consistent_assignments}.

An advantage of our voting scheme is that a single disambiguating example often suffices
to induce a correctly generalized parser. For example, if \slime trained on 10 square images and 1 non-square image,
our algorithm would regenerate a parser that works on rectangular images.
Indeed, in practice we have observed that
our voting scheme resolves the ambiguity in most cases.

\noindent\textbf{Robustness.}
A key insight is that, due to the downstream parser decision tree
(Section \ref{sec:parser}), \slime does not need rewrites to be correct for
all, or even many of the inputs: it only needs it to work for at least
\textit{one} of the inputs given any arbitrary set of inputs (including
recursively on the remaining inputs)\footnote{We do prefer to create a
parser that accepts as many inputs as possible.}. Our voting scheme guarantees this
to be the case.

\elcheaposubsection{Replace remaining constants with header bytes}
\label{sec:replace_constants}
The second step in this rewrite phase is to identify the remaining constants
that can be replaced with header bytes \eg \code{LOOP\_BOUND\_A}/\code{B} and
\code{MIN\_Y} in Figure \ref{fig:parser_61x76_BGR}. The header contains
metadata, such as the sizes of payloads in MT76x0 firmware containers, or the width and height of a bitmap.

\noindent\textbf{Header section identification.}
The header is defined as the section from the beginning of the file, up to the start of the first data chunk. Some file formats (\eg MT76x0 containers) contain a fixed-size header, followed by the data chunks. Other formats (\eg BMP) store the header size (or, equivalently, offset of the data chunk) as a field inside the header itself; this can be viewed as a small, fixed-size header that contains the size of additional header sections, thus reducing the problem to that of fixed-size headers.

Since \slime is not given the file specification, \slime must infer the header size. If \slime overestimates the header size, the search procedure will take longer, and there may be many false positives of header bytes that coincidentally match constants in the generated IR. Conversely, if \slime underestimates the header size, \slime may fail to replace some constants with header bytes. In both cases, \slime will still be able to parse 100\% of the training set (see ``Robustness'' subsection below), though the regenerated code may not fully generalize. \slime therefore starts with a small header size estimate (32 bytes), and increases it until \slime can generate a generalized parser.

\noindent\textbf{Rewriting process.}
\slime first filters out any
input files that are not consistent with the rewrites from the previous step
(\ie files that contradict the n-tuple chosen during voting).
Next, for each \code{LOOP\_BOUND} and \code{MIN\_Y} variable, \slime packs the concrete
value into different types: 8-bit, 16-bit, 32-bit, and 64-bit integers (little
endian), with both unsigned and signed variants (\eg the bitmap height
is negative to denote top-down bitmaps). \slime compares
the packed value against the header to find possible matching locations.
\slime also considers whether a concrete value can be rewritten as
the product (or negated product) of two header fields\footnote{The previous
step, ``Rewriting constants as expressions'' (Section
\ref{sec:rewrite_constants}) cannot rewrite the loop bound of top-down unpadded
bitmaps as a product of the width and height (and stride), because the width and
height do not appear as other constants, and there are generally many different
combinations of width/height that result in the same input/output index mapping.
For example, 1x4, 2x2, and 4x1 top-down BGRA bitmaps all share the same mapping,
hence it is impossible to infer the width/height from the mapping alone, without
comparing against the header bytes.}, or if two data chunks are adjacent (\eg \code{MIN\_Y} of one object is equal to \code{MIN\_Y} + \code{LOOP\_BOUND\_A} of another object).\footnote{In the general case, the formula needs to take into account strides, nested loops, and bottom-up mappings.}

For each variable, if there is no matching location, \slime assumes that it is a bona fide constant, and does not replace it. For example, MT76x0 firmware images have a fixed header size of 32 bytes, and therefore the first data chunk begins at byte 32; 32 bytes is a property of the file format, and does not appear as a field in the input file. Similarly, when \stbimage reads in
a 16-bit bitmap, it outputs a 24-bit image; 24-bit is a property of \stbimage, and
does not appear in the input file.

\noindent\textbf{Ambiguity.}
\slime cannot distinguish between signed \vs
unsigned if the input files do not have any values large enough to only fit in
unsigned integers, nor between different widths (\eg int16 \vs int32) if there are
coincidental zero values next to the variable (for example, \texttt{AD DE 00 00}
could be a 16-bit little-endian field [with value 0xDEAD] and 16-bits of zeros
to the right --- that may be padding or belong to another variable --- or a
32-bit little-endian field [with value 0xDEAD]).
Additionally, \slime cannot accurately identify the dimension header fields (\eg width and height for bitmap images) if the training data is ambiguous (\eg square bitmaps).

\noindent\textbf{Robustness.}
As per rewriting constant definitions (Section
\ref{sec:rewrite_constants}), \slime's downstream parser decision tree means that this step only needs to replace constants in a manner that works for at least \emph{one} of the input files.
\slime employs a similar n-fold Cartesian product voting scheme in this step.
Ties are broken by choosing the widest type possible. This allows generalization
to larger files, even when trained on small files. If \slime had erroneously
chosen a wider type, it would discover this relatively quickly, because any
counterexample input must be small. For example, suppose that the training data does not disambiguate whether the length field is 8-bit, 16-bit or 32-bit. If \slime chooses the widest type (32-bit), but the ground truth is that the length field is 8-bit, then there exists a counter-example input with a $\leq$ 255-byte data chunk.

\section{Merge parsers into decision tree}
\label{sec:parser}

In the worst case, the
algorithms so far are guaranteed to generate a parser -- albeit
possibly with concrete loop bounds --- that
can parse at least \textit{one} of the inputs from
the training set.
In this section, we explain how, using only this worst-case guarantee, \slime can
build a parser decision tree that converges to parsing all types represented in the training data.

\elcheaposubsection{Parse by Type}
\label{sec:parse-by-type}

\begin{definition}
A file's \textit{signature} corresponds to a subset of bytes in the
header that determine the execution path through the original parsing program.
We say two files share the same signature, if they have the same values
at the corresponding header positions.
\end{definition}
We consider any files with the same \textit{signature} to be of the same
\textit{type}. Files of the same type should be accepted by the same parser,
assuming the proper loop bound generalization and excluding files with
non-header-dependent data transformations. The
intuition is that we can model individual parsers, which generalize to
properties such as width and height (for bitmaps) or number of samples (for audio), for each type of input (\eg 16- \vs 24-bit
bitmaps, 8- or 16-bit audio), and then assemble these parsers into a decision tree that dispatches a
new input to the correct parser based on certain values in the input's header
(\eg the file signature).

We define a language (\Cref{fig:tree-dsl}) for constructing the parser
decision tree as a collection of individual parsers guarded by
predicates over file signatures. Nodes in this tree correspond to a predicate,
with a branch for input files that satisfy the predicate and a branch for those
that do not. A leaf in the tree corresponds to a parser we have regenerated. Parsing a file
corresponds to traversing this tree until a leaf is reached. The parser decision tree is generated as \slime IR, and
an appropriate backend handles code generation (Section \ref{sec:backend}).

\begin{figure}
\[
\begin{array}{rcl}
  \textit{tree} & := & \Call{Leaf}{\text{indivParser} \in \textit{Parsers} \vert null } \\
    & \vert & \Call{Node}{p \in \text{pred}, b_{\text{true}} \in \text{\textit{tree}}, b_{\text{false}} \in \text{\textit{tree}}} \\

\textit{pred} & := & \text{byte}_i = c \in \mathsf{Z}
\end{array}
\]
\caption{Tree-based parser DSL.
\textit{Parsers} corresponds to the set
of type-specific individual parsers regenerated so far in the \slime{} pipeline,
each of which can parse a type of input,
and \textit{pred} is the set of boolean predicates
over inputs' signatures.
}
\label{fig:tree-dsl}
\end{figure}

\elcheaposubsection{Building a decision tree, given a fixed set of \diode logs}
We begin with the simple case where we have a fixed set of \diode logs,
corresponding to a subset of input files.
Algorithm \ref{algo:tree-learning} presents the core algorithm to model and regenerate a
tree-based parser.
Given a newly generated parser, each input file can be annotated as
correctly or incorrectly parsed by that parser, by comparing the output of the
generated parser against the reference application (which plays the role
of a functionality oracle). If all examples have been
correctly parsed,
then we create a leaf that contains that parsing program,  \textit{indivParser}.

Section \ref{sec:identify_header}'s robustness guarantee is that for
any set of input files and their corresponding \diode logs, the
generated  \textit{indivParser} will parse at least one of the input files. Thus,
the only circumstance in which  \textit{indivParser} cannot parse any input files
is if we do not have any \diode logs. In this case, we create a leaf
with a null parser, where no files will be parsed.
This will be resolved when we expand our set of
\diode logs (discussed below).
If we have \diode
logs for all input files, we are guaranteed to produce a tree with leaves where
all inputs can be parsed, since we can trivially add a file's full signature
(one byte at a time) as the required predicate nodes in the tree.

If \textit{indivParser} works on only some of the inputs (\ie it fails
to parse a proper subset of the input files), we would ideally split
the inputs based on which ones are parseable, and then call the tree-building
algorithm recursively on only the unparseable inputs. We approximate
this behavior using the \code{pickAHew} function (Algorithm \ref{algo:pick-a-choose})
which finds
an equality predicate on a file signature byte that maximally
distinguishes the parseable \vs unparseable bitmaps.

\begin{algorithm}
\caption{Inferring a tree-based parser}
\label{algo:tree-learning}
\begin{algorithmic}
\Require{example input files $\mathit{examples}$, \diode log files $\mathit{logs}$, the original
application $\mathit{oracle}$}
\Ensure{A tree that can be compiled into the complete parser}

\Function{BuildTree}{$\mathit{examples}$, $\mathit{logs}$, $\mathit{oracle}$}
    \State $\mathit{indivParser} \gets \Call{BuildIndivParser}{\mathit{examples}, \mathit{logs}}$
    \State $(\mathit{parseable}, \mathit{unparseable}) \gets \Call{testParser}{\mathit{parseTree}, \mathit{corpus}, \mathit{oracle}}$

\If{$|\mathit{unparseable}| == 0$}
    \State \Return \Call{Leaf}{$\mathit{indivParser}$}
\ElsIf{$|\mathit{parseable}| == 0$}
    \State \Return \Call{Leaf}{$\mathit{null}$}
\Else
    \State $\mathit{predicate} \gets \Call{PickAHew}{\mathit{parseable}, \mathit{unparseable}}$
    \State $(\mathit{sat}, \mathit{unsat}) \gets \text{Split examples on } \mathit{predicate}$

    \If{$(|\mathit{sat}| == 0) \lor (|\mathit{unsat}| == 0)$}
        \State \Return $\Call{Error}{\text{"Cannot find predicate in header. Try increasing header size."}}$ %
    \Else
        \State $\mathit{left}  \gets \Call{BuildTree}{\mathit{sat}, \mathit{logs}, \mathit{oracle}}$
        \State $\mathit{right} \gets \Call{BuildTree}{\mathit{unsat}, \mathit{logs}, \mathit{oracle}}$
        \State \Return $\Call{Node}{\mathit{predicate}, \mathit{left}, \mathit{right}}$
    \EndIf
\EndIf
\EndFunction
\end{algorithmic}
\end{algorithm}

\begin{algorithm}
\caption{Finding an equality predicate, to create a split node
in the decision tree, that hews good \vs bad files
}
\label{algo:pick-a-choose}
\begin{algorithmic}
\Require{
$\mathit{goodFiles}$, which are
successfully parsed, and $\mathit{badFiles}$, which
fail to be parsed
}
\Ensure{An equality predicate over byte index $i$ and value $v$,
to hew files.}

\Function{PickAHew}{$\mathit{goodFiles}$, $\mathit{badFiles}$}
    \ForAll{$i \in [0 .. \mathit{headerSize})$}
        \ForAll{$\mathit{file} \in \mathit{goodFiles}$}
            \State $\mathit{freq} [i][\mathit{file} [i]] \pluseq \frac{1}{|\mathit{goodFiles}|}$
        \EndFor
        \ForAll{$\mathit{file} \in \mathit{badFiles}$}
            \State $\mathit{freq} [i][\mathit{file} [i]] \minuseq \frac{1}{|\mathit{badFiles}|}$
        \EndFor
    \EndFor

    \State $\hat{i}, \hat{v} \gets \argmax_{i,v} (|\mathit{freq} [i][v]| - i\epsilon)$ \Comment{$i\epsilon$ is a tie-breaker}
    \State \Return $\lambda f: \mathit{file}. f[\hat{i}] == \hat{v}$ \Comment{given a file, check equality of byte value}
\EndFunction
\end{algorithmic}
\end{algorithm}

\elcheaposubsection{Choosing a small set of logs}
\label{subsec:active-learning}
Applying Algorithm \ref{algo:tree-learning} to a complete set of \diode logs
would create a decision tree that parses the entire training set. However,
generating the necessary logs can be computationally
expensive, and wasteful, as
\slime can generalize from a few representatives of each input type.
Small inputs and large inputs of the same type create identical parsers (when appropriately generalized),
despite their different overheads; thus,
our goal is to generate logs only by executing the instrumented application with the smaller
input files.\footnote{Sometimes \diode logs from smaller inputs may have more
ambiguity when rewriting or replacing constants. If this results in a parser
that is not sufficiently general, Algorithm \ref{algo:active-learning-diode}
will correct this in a later iteration by generating logs for the larger,
unparseable input.}

Algorithm \ref{algo:active-learning-diode} shows our iterative process to select
the smallest unparseable input files, create \diode logs for these and
build the parser tree. The process repeats until the parser tree has sufficiently
high coverage of available inputs. We can inexpensively test whether an input is parseable by our current parser tree, by comparing its output with
that of the original application.

Our \slime{} prototype generates
logs for 10 unparseable files at a time to exploit CPU parallelism.\footnote{10 files is based on our workstation properties.}
If, at each iteration, we obtained the log for only the smallest unparseable file, we would end up with the minimal set of logs.

\begin{algorithm}
\caption{Choosing a small set of logs}
\label{algo:active-learning-diode}
\begin{algorithmic}
\Require{application $\mathit{app}$, input files $\mathit{corpus}$, \diode logs $\mathit{logs}$}
\Ensure{A tree that can be compiled into the full parser, and a small set of \diode logs}

\Function{expandLogsUntilConverged}{$\mathit{corpus}$, $\mathit{oracle}$}
    \State $\mathit{logs} \gets \mathit{nil}$, $\mathit{parserTree} \gets \mathit{null}$, $\mathit{parseable} \gets \mathit{nil}$, $\mathit{unparseable} \gets \mathit{corpus}$

    \While{($|\mathit{unparseable}| \geq 0$)}\Comment{coverage guarantee}
        \State $\mathit{fails} \gets \Call{getSmallest}{10, \mathit{unparseable}}$ \Comment{10 is based on workstation constraints}
        \State $\mathit{logs} \gets \mathit{logs} + \Call{GetDIODELogs}{\mathit{fails}}$
        \State $\mathit{parserTree} \gets \Call{buildTree}{\mathit{corpus}, \mathit{logs}, \mathit{oracle}};$
        \State $(\mathit{parseable}, \mathit{unparseable}) \gets \Call{testParser}{\mathit{parserTree}, \mathit{corpus}, \mathit{oracle}}$
    \EndWhile
    \State \Return $\mathit{parserTree}, \mathit{logs}$
\EndFunction
\end{algorithmic}
\end{algorithm}

\section{Emit Hardened Code}
\label{sec:backend}

\slime's use of an intermediate representation means that, by choosing a different backend, \slime can regenerate parsers in multiple languages: the same language as the original, for a drop-in replacement; or a different language, for automatic porting. \slime's prototype has two backends (C and Perl), which illustrate these two use cases, as well as vastly different points in the design space: C is a compiled language (Perl is interpreted); C lacks memory safety (\slime's wrapper functions add spatial memory safety in the regenerated parsers; Perl natively provides memory safety); and, critically, C directly exposes machine data types (e.g., \code{uint64\_t}; Perl does not).

\subsection{Backend \#1: Lowering Z3 to C}

Z3 code supports arbitrary-width bit-vectors~\cite{z3tutorial}. In the general case of converting Z3 to C, \slime would need to implement bit-vector support as well.\footnote{We could import the Z3 C bindings instead, but this adds a heavyweight dependency to the regenerated parser.} However, the key insight is that \diode's Z3 bit-vectors are used solely to represent/``lift'' LLVM IR, and hence are exactly 8/16/32/64/128-bit. \slime is therefore able to optimize the translation by mapping/``lowering'' these bit-vectors back to C's fixed-width data types (\eg \code{(\_ BitVec n)} maps to \code{int64\_t}/\code{uint64\_t})\footnote{Z3 does not have ``signed'' vs ``unsigned'' bitvectors; rather, it is necessary to choose the appropriate arithmetic operator.~\cite{z3tutorial}}); moreover, operations on Z3 bit-vectors map cleanly to operations using C's data types (\eg Z3's explicit zero-extend and sign-extend operators are equivalent to C code that assigns to a larger-width unsigned and signed int respectively; see  Table \ref{tab:lowering}).

\begin{figure}[!h]
\begin{tabular}{ | l | l | l | }
\hline
\textbf{Operation} & \textbf{Z3} & \textbf{C} \\
\hline
Sign extend  & (\_ sign\_extend 32) x & int64\_t y = (int32\_t) x; \\
Zero extend  & (\_ zero\_extend 32) x & int64\_t y = (uint32\_t) x; \\
Extract bits & (\_ extract 31 16) x    & int16\_t y = x >> 16; \\
\hline
Arithmetic shift-right & bvashr x bv16       & int32\_t y = ((int32\_t) x) >> 16; // sign-extend \\
Logical shift-right & bvlshr x bv16       & int32\_t y = ((uint32\_t) x) >> 16; // zero-extend \\
\hline
Shift-left          & bvshl x bv16        & int32\_t y = x << 16; // arithmetic shift = logical shift \\
\hline
\end{tabular}
\caption{Examples of lowering DIODE's Z3 expressions to C. Assume that \code{x} is a 32-bit bit-vector in Z3, declared as \code{int32\_t x;} in C.}
\label{tab:lowering}
\end{figure}

\subsection{Backend \#2: Emulating Z3 in Perl}

Figure \ref{tab:lowering} showed that C's fixed-width data types were essential for efficiently lowering Z3. Unfortunately, Perl does not directly expose machine data types; the interpreter can choose to store numbers as native integers, native floating-point, or decimal strings, converting between them as necessary~\cite{perlnumber}. The importance of explicit variable widths can be seen through a simple example. Suppose \code{x = -1}, and \code{x} is then zero-extended to a 16-bit bit-vector; it can be either: -1 (bit-vector: \code {1111111111111111}) if \code{x} was originally 16-bit, or 255 (bit-vector: \code{0000000011111111}) if \code{x} was originally 8-bit. Additionally, Perl does not expose all the primitives needed to efficiently map Z3's operators; for example, all bit-shifts are logical shifts by default, unless the ``use integer'' pragma is enabled, in which case all bit-shits become arithmetic shifts~\cite{perlop}.

\slime works around these limitations by ``emulating'', instead of lowering, Z3. For example, Z3's \code{bvmul} operator is mapped to a helper function \code{SLIMESupport::bvmul}, which internally uses the library function \code{Bit::Vector::Multiply}. The extra indirection of the \code{SLIMESupport} wrappers (instead of directly emitting \code{Bit::Vector::Multiply}) makes the code cleaner, and allows substituting the \code{SLIMESupport} library with a more efficient implementation in the future.

\section{Evaluation Methodology}
\label{sec:evaluation}

We implemented \slime and used it to empirically evaluate the effectiveness of \slime's algorithms. We describe the evaluation methodology in this section, and present the results in Section \ref{sec:results}.

\elcheaposubsection{Input Formats and Reference Parsers}
We evaluate \slime using the WAV (waveform audio), OS/2 1.x bitmap, Windows 3.x bitmap, Windows 95/NT4 bitmap, and Windows 98/2000 bitmap, and MT76x0 firmware container formats. WAV and the BMP family are classic, eponymous file formats for waveform audio and bitmap images files, respectively~\cite{bmp4_spec,bmp5_spec,wave_format}. Note that, as early as 1994 (\ie predating the Windows 95/NT4 and 98/2000 bitmap formats), the BMP family was already recognized as \emph{``not a standard file format ... the BMP format is actually a sheaf of formats bundled under the same name''}~\cite{bmp_dobbs_sheaf}. MT76x0 firmware containers are used to store part of the firmware updates for wireless routers that use the MediaTek MT76x0 chipset.

\slime's algorithms do not depend on how the mapping between input file bytes to data structure contents is obtained. For our prototype, we obtained these mappings by sending input files through three reference parsers, each of which has been instrumented with \diode (with some minor changes to work around limitations of the \diode prototype): \sndlib, \stbimage, and \openwrt~[\citeyear{openwrt}]\footnote{\url{https://github.com/openwrt/mt76/blob/master/mt76x0/usb_mcu.c}, with additional modifications to store the firmware container contents into a data structure rather than flashing the hardware device.}.
Table \ref{tab:formats} summarizes the file formats and reference parsers used in our evaluation.

\begin{figure}[!h]
\resizebox{0.82\columnwidth}{!}{
\begin{tabular}{ | l | l | l | l | }
\hline
\textbf{File Extension} & \textbf{File formats} & \makecell{\textbf{\# files in} \\ \textbf{corpus}} & \makecell{\textbf{Reference} \\ \textbf{Parser}} \\
\hline
WAV & RIFF (little-endian) data, WAVE audio & 1,651  & \sndlib \\
\hline
BIN & MT76x0 firmware containers & 5 & \openwrt \\
\hline
BMP & OS/2.1x format                        & 9      & \stbimage \\
    & Windows 3.x format                    & 9,971  & \\
    & Windows 95/NT4 and newer format       & 133    & \\
    &  Windows 98/2000 and newer format     & 895    & \\
    \cline{2-3}
    & \emph{Combined corpus (input to \slime)} & 11,008 & \\
\hline
\end{tabular}
}
\captionof{table}{File formats in our evaluation.}
\label{tab:formats}
\end{figure}

\elcheaposubsection{Corpora}

\noindent\textbf{WAV.} We used a compilation of ``999 WAVs for Windows''\footnote{\url{https://archive.org/details/cdrom-999wavsforwindows}}, which, name notwithstanding, contains 1585 WAV files. We excluded \code{YMMUD.WAV}, which is actually a plain-text file designed ``to allow for the random selection of configured wav files to be played''. To increase file diversity, we also included all 70 \code{.wav} audio files from the \code{C:\char`\\Windows\char`\\Media} folder of a Windows 10 system. \slime is not given the provenance of the WAV files; its input is the combined corpus of 1654 WAV files.

These files differ based on the number of channels (mono vs. stereo), sampling rate, and recording length. The mapping from input file to in-memory data structures can be non-trivial; for stereo audio files, the input is interleaved channels, each containing 16-bit \code{short}s, while the output is separate (left- and right- channel) arrays of 64-bit \code{long}s.
Additionally, three WAV files\footnote{\code{BIG.WAV}, \code{BUTTHEAD.WAV}, and \code{CAFE80S.WAV}} contain extra metadata sections (instead of the canonical format-plus-data layout~\cite{wave_pcm}); we did not remove these files from the corpus, and \slime is able to parse them.

\noindent\textbf{MT76x0 .BIN firmware containers.} We downloaded all 621 driver tarballs (\code{.tar.gz}) from \citet{mt7610_drivers}. These tarballs contained 1106 .BIN files, which represented 2 unique variants\footnote{\code{mt7610u.bin}, \code{mt7650u.bin}}; the tarballs contain changes to other files (\eg \code{.ko} Linux kernel drivers), which are not in scope for this file format. Additionally, the Open Wireless Router~\cite{openwrt} project's sample firmware directory contains a further three unique variants\footnote{\code{mt7610e.bin}, \code{mt7662.bin}, \code{mt7662\_firmware\_e3\_v1.7.bin}}, for a total corpus size of five firmware containers.

\noindent\textbf{BMP.}
\slime can regenerate parsers for each of the four bitmap file formats (OS/2.1x, Windows 3.x, Windows 95/NT4, Windows 98/2000), each of which contains various subtypes (\eg bottom-up \vs top-down; 16-bit vs. 24-bit, \etc). We will show a strictly stronger result: \slime can regenerate a parser decision tree that parses the combined corpus of all these file types.
Files range in size from 6,966 bytes (48x48 24-bit) to 36,578,358
bytes (3024x4023 24-bit), with a total corpus size of 16,283,146,489 bytes (11,008 files).

We built the corpus by obtaining the top 10 most popular search phrases for each of the 25 English-text categories from Google's
Year in Search 2018 \cite{google_yis_2018}.
We then used \code{google\_images\_download} \cite{google_images_download} to scrape the top 100 bitmap images from Google Image Search for
each of the 232 unique keywords.
This returned a median of 68 bitmaps per phrase (min 1, max 92) due to various errors
(\eg ``Wrong image format returned''), for a total of 13,694 bitmaps.
We filtered out: four files that are not ``PC bitmap'' (which were erroneously included by \code{google\_images\_download}), duplicate (byte-for-byte identical) copies of bitmaps (1980 files)\footnote{\slime can successfully handle duplicates, but it inflates the correctness rate.}, palettized bitmaps (448 files), and corrupted bitmaps (168 files which contains less pixel data than implied by the dimensions, and 86 32-bit bitmaps with all-zero alpha channels).

\elcheaposubsection{Training/test performance of entire parser tree} \slime regenerates the parser in the form of a decision tree, with individual parsers at the leaves. The first way to evaluate \slime's performance is to consider the overall correctness of the regenerated parser.
Algorithm \ref{algo:active-learning-diode} guarantees 100\% correctness on the training set. To evaluate the generalizability of our automatically generated WAV and BMP parsers, we performed standard
held-out cross-validation~\cite{isl}. For each of the WAV and BMP corpora, we randomly split 80\% into a training set
(1353 WAVs or 8806 BMPs), ran the complete \slime pipeline, and then tested on the
remaining held-out 20\%. We repeated this process a total of 100 times
to account for the random split.
For MT76x0 firmware containers, due to the smaller corpus size, we used a single firmware container as training input, and checked that \slime's generated parser could parse the other firmware containers.

\elcheaposubsection{Automatically identified predicates at nodes of parser decision tree}
The parser decision tree, as a whole, is useful for parsing the entire corpus. However, it is also notable that, for each of the nodes, Algorithm \ref{algo:pick-a-choose} can automatically choose predicates that are meaningful for the file format, \textit{without} using the file specifications, nor any explicit control-flow information from the original parser (recall that the \diode logs do not contain such information).
We applied the \slime pipeline to the complete WAV and BMP corpora to create decision trees that can parse all the input files, and discuss some of the predicates that \slime has chosen.
The MT76x0 firmware containers could all be parsed with a single leaf parser, hence there were no predicates identified nor needed.

\elcheaposubsection{Optimized processing}
We estimate the runtime speedup of \slime's \textsc{ExpandLogsUntilConverged} (Algorithm \ref{algo:active-learning-diode}) compared to other file selection strategies.

\elcheaposubsection{Bugs prevented or found}
We discuss how \slime's regeneration process automatically prevents memory-safety bugs.
Additionally, we will discuss two new bugs in the \stbimage library that we found and
reported\footnote{\url{https://github.com/nothings/stb/issues/773}, \url{https://github.com/nothings/stb/issues/783}}, and also a known bug that
we independently rediscovered. Neither of the new bugs lead to memory-safety
issues, and hence would not have been avoided by (re-)writing \stbimage in a memory-safe
language or running \stbimage with a memory error detector such as AddressSanitizer~\cite{address_sanitizer}.
We also conceptualize the general classes of bugs that \slime can help detect.

\section{Results}
\label{sec:results}

\elcheaposubsection{Training/test performance of entire parser tree}
\label{sec:evaluation_bmp}

\noindent\textbf{WAV parser.}
\slime's decision trees correctly parsed between 99.76\% and 100.00\% of the entire WAV corpus (mean: 99.98\%). The parsing failures occur when rare WAV classes are represented in the test set but not the training set.

\noindent\textbf{BMP parser.}
\slime's decision trees correctly parsed between 99.93\% and 100.00\% of
the entire bitmap corpus (mean: 99.98\% --- coincidentally the same as the WAV parser). The parsing failures occur when
rare BMP classes are represented in the test set but not the training set.
For comparison, the original parser correctly parses only $\approx 98\%$ of the
corpus, because it contains bugs which \slime helped us find and fix (Section \ref{sec:memory_safety}), and which our regenerated parser avoids.

\noindent\textbf{MT76x0 firmware container parser.}
Training on any single file from the corpus sufficed to create a generalized parser for all inputs.

\noindent\textbf{Porting to other languages.}
\slime's backends are able to regenerate working C or Perl code that implement the decision trees (including the leaf parsers).
Regenerating code in another language can be achieved by adding a different backend; \slime reuses the results from earlier steps (loop summarization, rewriting expressions, decision tree) in \slime's pipeline, as those are performed exclusively on the intermediate representation. This means that \slime can produce high-correctness parsers for other languages, for which the alternative may otherwise be no parser available.

\elcheaposubsection{Automatically identified predicates at nodes of parser decision tree}

\begin{figure}[!h]
\centering
\includegraphics[scale=0.66]{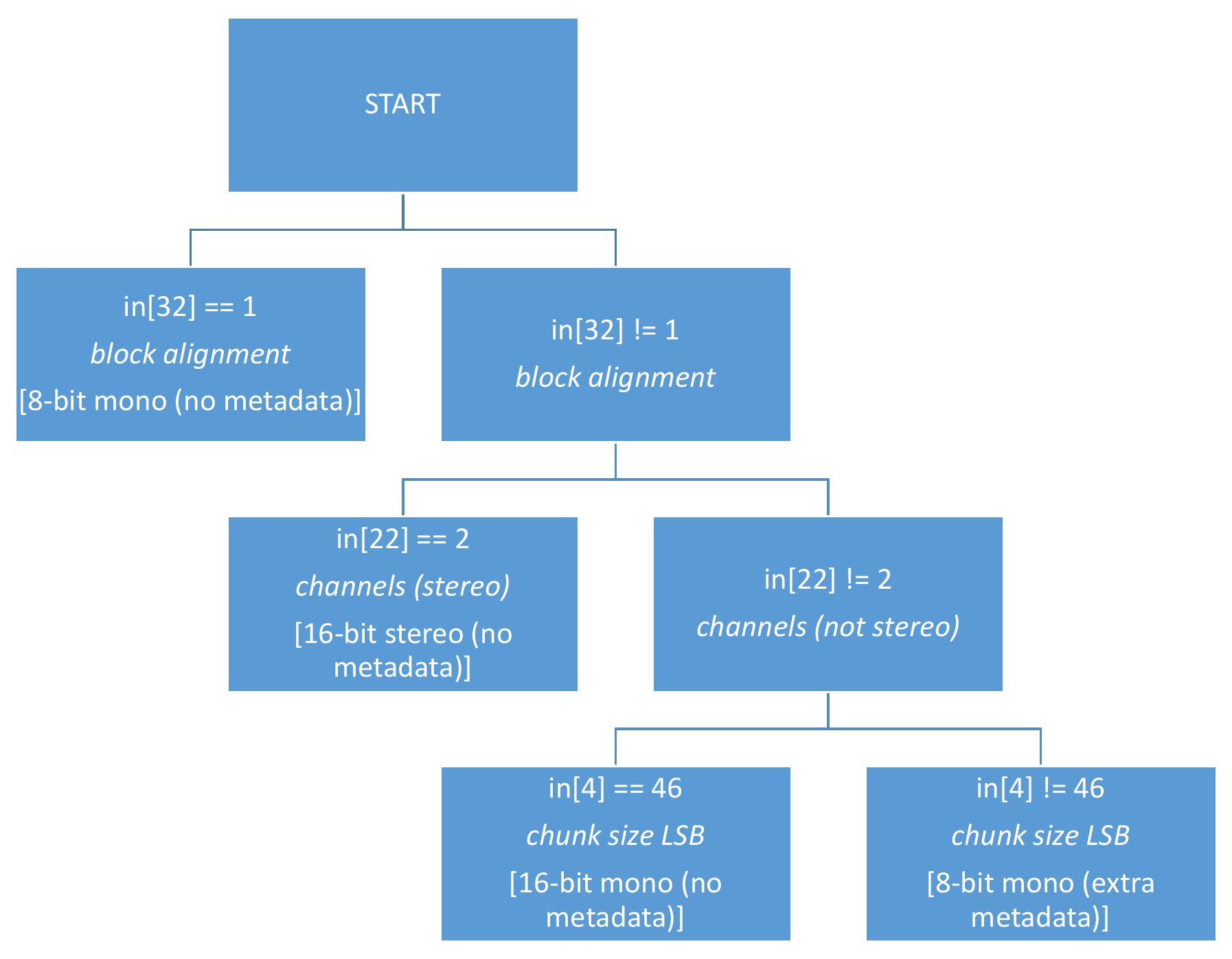}
\caption{
\textbf{Generated parser decision tree for \code{SndInfo}'s WAV functionality.}
The first line of each node denotes the decision tree predicate from its parent; the
second line relates that inferred predicate to a key file format property.
The last line of each leaf explains the file formats that can be parsed.
}
\label{fig:snd_info_parse_tree}
\end{figure}

\noindent\textbf{WAV parser.}
The generated WAV parser decision tree (Figure \ref{fig:snd_info_parse_tree}) is a complete binary tree with 4 leaves (each leaf corresponds to a parser for a specific input file format) that correctly parses all 1651 WAV files in the corpus (\ie 100\% correctness). At each node, we show the predicate from its parent. For example, the root is split based on whether byte 32 of the file --- the ``block alignment'' of the audio (\ie number of channels * bytes per sample) --- is 1, which corresponds to 8-bit mono audio; meanwhile, the predicate \code{in[22] == 2} checks the number of audio channels (\ie if it is stereo). The predicates are all automatically selected by \slime (Algorithm \ref{algo:pick-a-choose}), without using the WAV file specification.
The leaves describe each parser's functionality; for example, the left-most leaf corresponds to a parser for 8-bit mono audio files that lack extra metadata.

\begin{figure}
\centering
\includegraphics[scale=0.82]{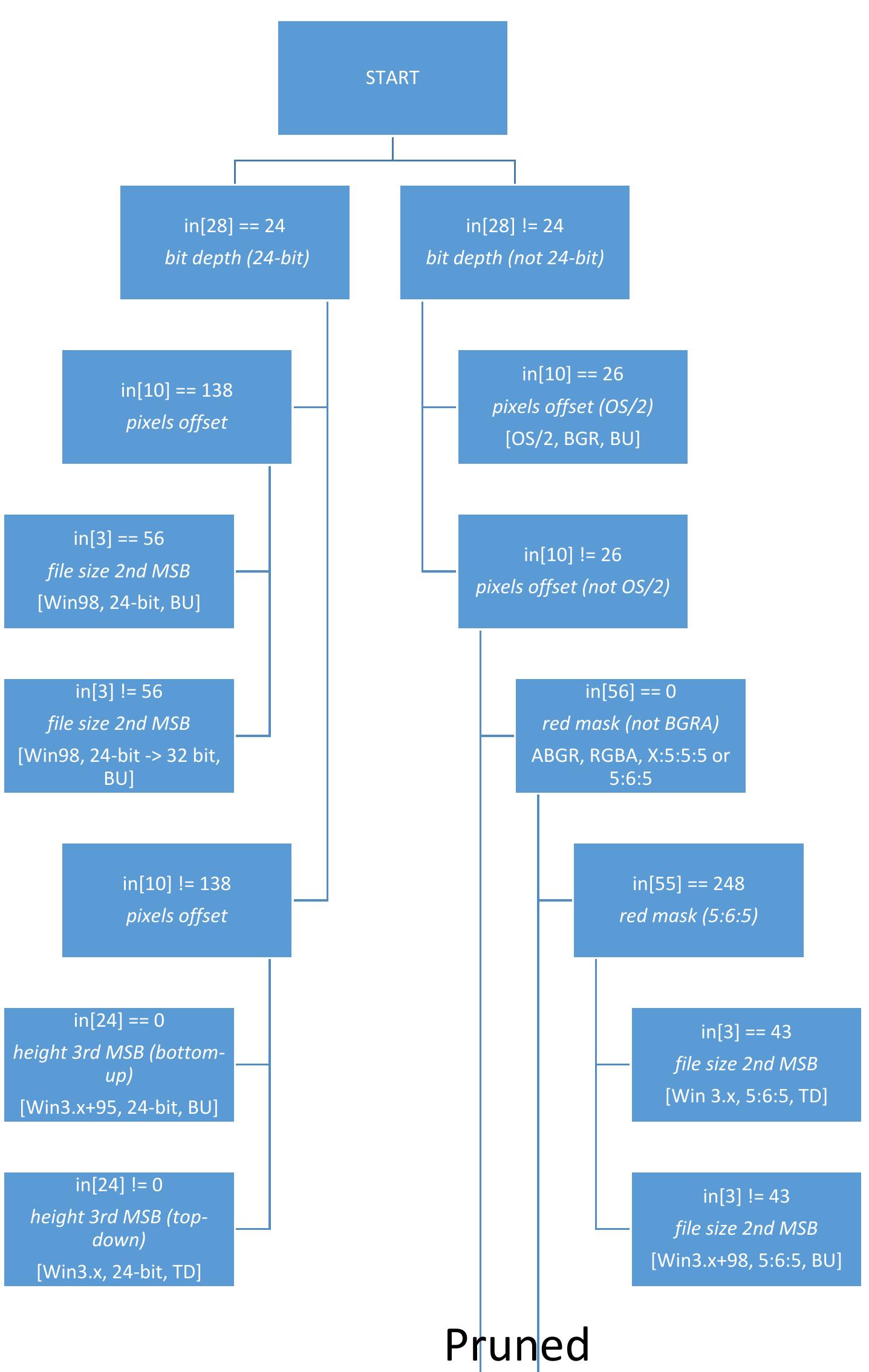}
\caption{
\textbf{Generated parser decision tree for \stbimage's BMP functionality.
}
The first line of each node denotes the decision tree predicate from its parent; the second line relates that inferred predicate to a key file format property. The last line of each leaf explains the file formats that can be parsed.
}
\label{fig:stb_image_parse_tree}
\end{figure}

\noindent\textbf{BMP parser.}
The generated BMP parser decision tree (Figure \ref{fig:stb_image_parse_tree}) is a complete binary tree with 20 leaves that correctly parses all 11,008 bitmaps (\ie 100\% correctness). \slime's Algorithm \ref{algo:pick-a-choose} once again automatically selects relevant predicates, without using the file specification. For example, the root is split based on whether byte 28 of the file --- the bit-depth of the image --- is 24. The predicate \code{in[10] == 138} checks the offset of the pixel data in the file, or equivalently, the size of the header and therefore the ``version'' of the bitmap file format. Another internal node identifies whether an image is top-down (\code{height < 0}) or bottom-up (\code{height > 0}). Although our predicate selection (Algorithm~\ref{algo:pick-a-choose}) is designed to test strict equality of individual bytes, it is able to determine if the height field is negative by taking advantage of the two's complement representation of 32-bit integers (\code{in[24] == 0}). Additionally, while our predicate selection algorithm is designed to test single bytes, it chooses the correct parser for different 4-byte bitmasks by recursively partitioning on individual bytes until they are disambiguated, in a similar fashion to Blind ROP's byte-by-byte discovery of stack canaries~\cite{bittau2014hacking}. Unlike Blind ROP's hardcoded strategy, this byte-by-byte discovery is an emergent property of \slime's decision tree algorithm.

\elcheaposubsection{Optimized processing}

Table \ref{fig:active_learning} shows the number of unparseable WAVs or BMPs (across all leaves of their parser decision tree) at the start of each
round when using \textsc{ExpandLogsUntilConverged} (Algorithm \ref{algo:active-learning-diode}).

The WAV parser results required two iterations of \textsc{ExpandLogsUntilConverged}, with \diode logs for only 20 of the 1,654 WAVs. At the start of round 0, the entire corpus (1,654 WAVs, totaling 121MB) cannot be parsed. The smallest 10 WAVs total just 12KB; after obtaining \diode logs for these smallest WAVs, only 73 WAVs ($5\%$ of the corpus by file count; $16.5\%$ by file size) are unparseable. We then repeat the process by selecting the 10 next smallest unparseable WAVs (377KB), which suffices to reach convergence.
In total, \slime required \diode logs for just 0.4MB of WAVs (0.32\% of the corpus by file size).
The BMP parser results required eight iterations of \textsc{ExpandLogsUntilConverged}, with \diode logs for only 74 of the 11,008 BMPs. In total, \slime obtained \diode logs for just 12.8MB of BMPs (0.08\% of the corpus by file size).

\begin{figure}[!h]
\resizebox{0.92\columnwidth}{!}{
\begin{tabular}{ | c | r | r | r | r | r | r | r }
\hline
& \multicolumn{3}{c|}{\textbf{Unparseable WAVs}} & \multicolumn{3}{c|}{\textbf{Unparseable BMPs}} \\
\hline
Round & \makecell{\# files} & \makecell{\# bytes (all)} & \makecell{\# bytes \\ (10 smallest)} & \makecell{\# files} & \makecell{\# bytes (all)} & \makecell{\# bytes \\ (10 smallest)} \\
\hline
0     & 1654 & 121,464,191 & 11,747 & 11,008 & 16,283,146,489 &   131,492  \\
1     & 73   & 19,966,796 & 376,862 &  1,809 &  2,913,068,807 &   328,480  \\
2     & 0    & 0          & 0       &  1,260 &  2,059,278,713 &   591,480  \\
3     &      &            &         &  1,085 &  1,913,469,901 &   871,512 \\
4     &      &            &         &    308 &    736,009,768 & 1,247,716 \\
5     &      &            &         &    106 &    284,912,770 & 1,733,040 \\
6     &      &            &         &     58 &    191,820,812 & 3,890,752 \\
7     &      &            &         &      4 &      3,992,704 & 3,992,704 \\
8     &      &            &         &      0 &              0 &         0 \\
\hline
TOTAL &      &            & 388,609 &        &                & 12,787,176 \\
\hline
\end{tabular}
}
\captionof{table}{
\textbf{
Number of unparseable BMP or WAV files (across all leaves)
at different algorithm iterations.
}
Round 0 is the initial corpus of 1654 WAVs or 11,008 BMPs. At each round, \slime obtains \diode logs for the 10 smallest unparseable bitmaps, and then re-runs
\code{BuildTree} (Algorithm \ref{algo:tree-learning}).
}
\label{fig:active_learning}
\end{figure}

\noindent\textbf{Comparison with other file selection strategies.}
We estimate that \slime's optimized processing is orders of magnitude faster than other strategies\footnote{There are corner cases where it is not optimal. For example, suppose that the corpus consists of 10 small files of the same type, plus a large file of a different type: processing the 10 smallest files first would be less efficient than processing the 10 largest files.}. Since processing using those other strategies can take hundreds, or even thousands of CPU days to complete, it is impractical to directly measure their runtime. Instead, we estimate their runtime by measuring the amount of time required to generate
 \diode logs
for a subset of WAV and BMP files, then using these measurements to build performance models that characterize the time required to obtain the final \slime parser. With these models, the time to process a WAV file is $0.2157B$, and the time required to process a BMP file is
$0.0692B$ seconds, where $B$ is the number of bytes in the WAV or BMP file (\eg a 10,000 byte BMP file takes approximately 692 seconds to process).
Table \ref{fig:log_strategies} uses these models to compare the number of bytes of input needed to be run through the instrumented parser, and corresponding approximate CPU days required, for various file selection strategies.

In the absence of Algorithm \ref{algo:active-learning-diode} --- which used the insight that we can test whether inputs are already parseable by our regenerated parsers --- we would need to generate \diode logs for all the files and feed them into Algorithm \ref{algo:tree-learning}. For the bitmap corpus, this would be prohibitively expensive: roughly 13,000 CPU days. This is orders of magnitude slower than Algorithm \ref{algo:active-learning-diode}, which only requires around 10 CPU days. We do have some extra overhead from rebuilding the tree and testing the parsers, but this
is negligible compared to \diode's runtime.

We also model the runtime of Algorithm \ref{algo:active-learning-diode} with a different heuristic than choosing the smallest unparseable files at each round. When choosing the \textit{largest} unparseable files, the modeled runtime is 500 CPU days; this is still much faster than running \diode on all the logs, but also significantly slower than choosing the smallest files. We also tried random selection: the modeled runtimes were in between (roughly 70-80 days).

The results for the WAV corpus follow the same general pattern as the BMP corpus, again showing that Algorithm \ref{algo:active-learning-diode} with our smallest size heuristic provides a significant speedup (1 CPU day instead of up to 300 CPU days).

\begin{figure*}[!h]
\centering
\resizebox{0.92\columnwidth}{!}{
\begin{tabular}{ | c | r | r | r | r | }
\hline
& \multicolumn{2}{c|}{\textbf{BMP}} & \multicolumn{2}{c|}{\textbf{WAV}} \\
\hline
\textbf{Strategy} & \makecell{\textbf{Input Bytes}} & \makecell{\textbf{Est. CPU Days}} & \makecell{\textbf{Input Bytes}} & \makecell{\textbf{Est. CPU Days}} \\
\hline
Process all files & 16,283,146,489 & 13,041.6 & 121,464,191 & 303.2 \\
\hline
Smallest size first & 9,388,502 & 7.5 & 388,609 & 0.97 \\
\hline
Largest size first & 621,022,028 & 497.4 & 10,399,639 & 26.0 \\
\hline
Random (seed 1) & 82,031,734 & 65.7 & 2,594,851 & 6.5 \\
Random (seed 2) & 78,058,706 & 62.5 & 3,385,886 & 8.5 \\
Random (seed 3) & 69,596,708 & 55.7 & 4,934,947 & 12.3 \\
\hline
\end{tabular}
}
\captionof{table}{
\textbf{
Estimated CPU days for different file selection strategies, based on our performance model.
}
}
\label{fig:log_strategies}
\end{figure*}

\elcheaposubsection{Bugs prevented or found}
\label{sec:memory_safety}

\noindent\textbf{Memory safety enhancements.}
Our IR examples (\code{DATA\_STRUCTURE [x] = ...}) are evocative of direct array accesses. \slime's C backend generates code that uses a wrapper function to prevent reading past the input bounds (Appendix \ref{sec:wrapper_read_input}), and replaces the output array with a dynamic array to prevent any buffer over/under-flows (Appendix \ref{sec:dynamic_array}). These enhancements are enabled in the backend, without needing developer input, changes to the original parser, or identification of any such bugs. \slime's Perl backend omits these checks, since Perl internally enforces them.

\noindent\textbf{New bug \#1 (not memory-safety related): missing bytes.} For 16-bit bitmaps with
compression type 3 (user-specified red/green/blue
bitmasks), \stbimage skipped the first 12 bytes of the pixel data, because
it did not account for the bitmasks when seeking from the end of the header to the start of the pixel data.
We were
able to identify this bug --- \textit{without} comparing \stbimage to a second parser ---
because the \diode logs for these
bitmaps were also missing the first 12 bytes, which made loop abstraction
impossible. %
The \stbimage maintainer noted the bug report and fixed it in a subsequent release.

\noindent\textbf{New bug \#2 (not memory-safety related): irrelevant bitmask.} For 32-bit bitmaps with
compression type 0 (no bitmasks), the user-specified
color channel bitmasks should be ignored according to the
specification~\cite{bmp4_spec,bmp5_spec}. \stbimage correctly checks the
compression field for Windows 3.x bitmaps, but for Windows 95/98/2000 bitmaps,
it always uses the bitmasks.
For the vast majority of Windows 95/98/2000 32-bit bitmaps, this does not result
in a bug, because the color channel bitmasks contain values equivalent to
the default (BGRA). However, for $\approx 1\%$ of our corpus, this has a
visible discrepancy, because the bitmask specifies RGBA. Our tree partitioning
algorithm grouped these images into their
own ``RGBA-parser'' leaf, making it easy to identify this class of bug-triggering
inputs by inspection.

\noindent\textbf{Known bug \#1: irrelevant bitmask redux.} For 24-bit bitmaps where the
``alpha mask'' header field is non-zero, \stbimage outputs a \emph{32-bit} (RGBA) image with the alpha channel set to 255 (fully opaque). Although the input and output images are visually identical, this is not the desired API behavior (as shown by the maintainer having fixed this bug), unnecessarily increases memory usage by one-third, and could cause a buffer overflow if the library user had only allocated enough for memory for a 24-bit image and blindly copied \stbimage's bloated output.
We identified this bug because our tree partitioning algorithm grouped the bug-triggering inputs (all but one of the
24-bit Windows 98 bitmaps) into their own leaf.
This bug has some overlap with our New bug 2; however, the fix for this bug (as of \stbimage 2.25) is highly specific
to 24-bit bitmaps and does not address the new bugs discussed previously.

\noindent\textbf{Classes of detectable bugs.}
\slime, by design, cannot automatically confirm that a behavior is buggy; strange behaviors can be deliberate ``features,'' and \slime's decision tree will faithfully replicate them. It does, however, assist with triaging the test cases for manual review, as files with the same behavior/bug will often be handled by the same leaf parser.

\slime can help detect bugs that manifest in buggy images being parsed by separate, buggy parser(s) (\eg in our new bug \#2, there was a leaf for Windows 95/98/2000 32-bit bitmaps with an RGBA bitmask). Conceptually, the decision tree has extra leaves for the buggy inputs, where, by some unusual criterion, the buggy inputs are separated from regular inputs. Debugging is simplified because we have both the test case, \emph{and} the criteria that distinguishes buggy \vs non-buggy inputs.

We can also quickly test the parser decision tree by inspecting the parser's output for one bitmap corresponding to each leaf. If the bug-triggering inputs are underrepresented in the corpus --- likely the case, as the bug would otherwise have already been found through ordinary testing --- then inspecting each leaf parser (with one test case each) is much more efficient than randomly sampling input files. The bug-triggering bitmaps for the two new bugs we identified are rare (each $\approx 1\%$ of our bitmap corpus),
highlighting both the coverage of our corpus and \slime's sensitivity to bugs.

Two special subclasses of detectable bugs are parsers that violate our high-level model of parsed data structures (that data is hyperrectangular, with dimensions defined by header fields; see new bug \#1) and buffer overflows (known bug \#1 may lead to this in application code, under some circumstances). These two subclasses result in parsing behavior that is so peculiar that they can often be manually confirmed as bugs, without comparing the output to another parser.

\slime does not assist with detecting bugs that manifest in buggy images being parsed by an inappropriate, but preexisting parser (\ie if the bug has resulted in merged or switched leaves/parsers in the decision tree). However, if these bugs result in buffer overflows, \slime will still prevent them by construction in the regenerated parser (as discussed earlier), even if we could not detect them in the original parser.

\section{Related Work}
\label{sec:related_work}

\subsection{Partial Input Language Specification}

Many papers~\cite{grimoire,veritas,prospex,discoverer,tupni} extract \emph{partial} input language specifications. Although these have many valuable use cases (\eg intrusion-detection systems, faster fuzzing), they are insufficient for \slime's use case of regenerating parsers that create byte-for-byte identical data structures. For example, in the context of BMP images, a partial specification would roughly be:

\begin{lstlisting}[basicstyle=\blobformat, escapeinside={<@}{@>}]
    width  : offset : length field;
    height : offset : length field
    bit-depth : offset : integer (with some constraints)
    pixels : offset : raw data (length = width * height * bit depth)
\end{lstlisting}

Notably, the \code{pixels} field is treated as raw bytes. This is adequate for many fuzzing use cases; the fuzzer would then fill in this field with random byte values to maximize
path coverage or similar metrics. Similarly, an intrusion-detection system could use these partial specifications to identify BMP files.
However, unlike \slime, a ``parser'' generated from this partial specification does not account for the different types of bitmaps, which each interpret the pixels differently (e.g., 16-bit, 24-bit, 32-bit with various bitmasks, top-down vs. bottom-up, etc.), let alone be able to generate working C or Perl code that can replace the original BMP library, which \slime can.

Additionally, it is worth noting that even a complete \emph{file format} specification (\eg BMP) is not generally sufficient to regenerate drop-in replacement libraries. The file format specification describes the \emph{input} format to the library, but the \emph{output} format of a library is library-dependent. For example, \stbimage converts all bitmap images to a top-down representation, even though BMP files are mostly bottom-up~\cite{bmp_dobbs}. However, BMP files are largely stored bottom-up because it is the native representation for Windows APIs, which use a Cartesian-coordinate system; thus, some parsers may prefer to output a data structure that is bottom-up. Similarly, while \sndlib's \code{mus\_file\_to\_float\_array} function returns all the samples from a single audio channel (\eg left or right), another library may prefer to keep the original WAV data format, which interleaves samples from the left and right audio channels (L0, R0, L1, R1, ...). \slime's modeling process uses the input/output mappings for the target library, thus its regenerated parsers produce data structures that are byte-for-byte identical to those of the original parsers.

\subsection{Language and Program Inference}
Prior work has extensively studied inference of languages, such as regular and
context free languages ~\cite{angluin1987learning, angluin1990negative,
de2010grammatical, gold1978complexity}. \citet{bastani2017synthesizing} apply a blackbox approach to inferring
programming languages by executing the original program. \citet{hoschele2016mining} use dynamic taint analysis to infer a context
free grammar for different input languages. In contrast to these, \slime{}
regenerates a parser for binary input formats with non-context-free features
such as header-dependent variable-length data segments.
Furthermore, \slime{}'s regenerated program does not just accept an
input, but also populates the original output data structures.

~\citet{caballero2013automatic}, \citet{caballero2007polyglot}, and \citet{caballero2009dispatcher} propose algorithms (including
dynamic-analysis based techniques) to infer the message structure of network
protocols. \slime{} also infers the structure of binary formats, but
extends this to handle more complex features, such as different WAV/BMP types and
variable-length segments.

~\citet{mining_input_grammars} learn context-free grammars by tracking accesses to the input buffer, as well as the control-flow of the original program, which is assumed to be a stack-based recursive-descent parser. \slime's inferrable file formats (Section \ref{sec:in_scope}) are not context-free. Furthermore, \slime does not use any information about the control-flow information of the original program, nor assume that it is structured in a particular way (as long as it parses the file format correctly).

Program inference shares characteristics with program synthesis, particularly
inductive program synthesis (aka \allowbreak programming-by-example)~\citep{polozov2015flashmeta, singh2016blinkfill, raza2018disjunctive}, which
uses example inputs and outputs as a partial program specification. In contrast,
\slime{} assumes we have access to the original application as a form
of specification, to produce new executions and rule out
incomplete/incorrect parsers.

Helium~\citep{mendis2015helium} learns in-memory ``stencil kernel'' transformations (such as Photoshop's
blur filter) and regenerates them as Halide DSL code, with the aim of higher
performance. \slime targets input processing code that parses a file into data structure contents, focusing on parser correctness and security.  Helium and \slime
solve some similar problems, such as ``buffer structure reconstruction'' (including stride detection) and canonicalizing expressions. Helium learns from a single example, while \slime integrates the information from multiple examples, with two key corollaries. First, Helium requires that the single input ``exercise both branches of all input-dependent conditionals.'' If there exists an input-dependent conditional for which no single input will exercise both branches (\eg in the context of parsers, two different file types), Helium cannot learn it; in contrast, \slime can model and regenerate parsers for each example, and unify them into a parser decision tree. Second, to learn input-dependent conditionals, Helium requires that its instrumentation record the branch taken/not-taken information. \slime does not; instead, \slime infers conditionals by comparing the header bytes between multiple examples.

Active learning has been used in machine learning extensively to improve
data efficiency~\cite{settles2009active}. \citet{alse} propose a
general framework that applies active learning and program inference to
learn and regenerate large applications.
Our work aligns with this framework, but we present detailed
algorithms and a concrete implementation of a prototype (\slime{}), along with
empirical evaluation.

Recent developments in program inference have demonstrated the potential for
learning non-trivial applications. For example, ~\citet{konure} infer programs that interact with a database by observing the
database query and result traffic. \citet{wu} used dynamic analysis to infer a
program's list/map functionality and replace these with a database. Similarly,
\slime{} aims to regenerate the full functionality of the original file parsers and does so using dynamic analysis, but \slime{} focuses on file parsing rather than database-related applications.

\subsection{Secure Parsers}
Manually constructed parsers are often
implemented as stateful shotgun parsers~\citep{momot2016seven}, where parsing
and input validation
functionality is scattered throughout an application, leading to subtle bugs.
This phenomenon has been observed in practice and can lead to vulnerabilities
~\citep{underwood2016search}. Prior work has developed tools to address these
difficulties. For example, Nail~\citep{bangert2014nail} is a tool to generate
secure data parsers based on a specification, and
Caradoc~\citep{endignoux2016caradoc} is a secure PDF parser and validator. \slime{}, in contrast
to Nail, does not require the user to write a specification to build a secure parser, but rather infers a parser from the original application's
executions and generates
safe-by-construction code. Caradoc's grammar and rules are hardcoded for PDF,
and there is no automatic process for inferring new formats from existing parsers.
In contrast to Caradoc, \slime{} has a modeling process and a simple set of primitives
that allow it to regenerate parsers for multiple file formats.

\section{Conclusion}
\label{sec:conclusion}

We presented \slime, the first system to model and regenerate a full working parser from instrumented program executions, producing data 
structures that are byte-for-byte identical to the data structures in the original program, by Summarizing Loops, Identifying header fields and 
expressions, Merging parsers into a decision tree, and Emitting hardened code.

Our empirical evaluation demonstrated that \slime can efficiently regenerate parsers --- while also preventing buffer overflows -- for six file 
formats (waveform audio, MT76x0 .BIN firmware containers, OS/2 1.x bitmap images, Windows 3.x bitmaps, Windows 95/NT4 bitmaps, and Windows 98/2000 
bitmaps), with a target language (C and Perl in our prototype backends\footnote{We leave JIT compilation --- \ie Just-in-time \slime --- for 
future work.}) that need not be the same as the original code. Additionally, \slime helped triage two new bugs, and its generated decision tree 
identified file format predicates that we found to be useful for identifying anomalies in the existing parser and for enhancing our understanding 
of the file format.

\section{Acknowledgements}

We thank the anonymous reviewers for their feedback, especially the reviewer who called the font and approach ``fresh''. Thank you to Shivam Handa for helpful discussions, Nikos Vasilakis for comments, Aarno Labs (particularly Ricardo Baratto and Eli Davis) for assistance with \diode, and Delaram Sadaghdar for the definition of GIF.

\newpage

\bibliographystyle{ACM-Reference-Format}
\bibliography{paper.bib}


\begin{thebibliography}{51}

%%% the Web address.

\ifx \showCODEN    \undefined \def \showCODEN     #1{\unskip}     \fi
\ifx \showDOI      \undefined \def \showDOI       #1{#1}\fi
\ifx \showISBNx    \undefined \def \showISBNx     #1{\unskip}     \fi
\ifx \showISBNxiii \undefined \def \showISBNxiii  #1{\unskip}     \fi
\ifx \showISSN     \undefined \def \showISSN      #1{\unskip}     \fi
\ifx \showLCCN     \undefined \def \showLCCN      #1{\unskip}     \fi
\ifx \shownote     \undefined \def \shownote      #1{#1}          \fi
\ifx \showarticletitle \undefined \def \showarticletitle #1{#1}   \fi
\ifx \showURL      \undefined \def \showURL       {\relax}        \fi
% invisible to TeX
\providecommand\bibfield[2]{#2}
\providecommand\bibinfo[2]{#2}
\providecommand\natexlab[1]{#1}
\providecommand\showeprint[2][]{arXiv:#2}

\bibitem[\protect\citeauthoryear{??}{bmp}{1994}]%
        {bmp_dobbs_sheaf}
 \bibinfo{year}{1994}\natexlab{}.
\newblock \bibinfo{title}{{The BMP File Format}}.
\newblock
\newblock
\urldef\tempurl%
\url{https://www.drdobbs.com/cpp/the-bmp-file-format/184409305}
\showURL{%
\tempurl}


\bibitem[\protect\citeauthoryear{??}{bmp}{1995}]%
        {bmp_dobbs}
 \bibinfo{year}{1995}\natexlab{}.
\newblock \bibinfo{title}{{The BMP File Format, Part 1}}.
\newblock
\newblock
\urldef\tempurl%
\url{https://www.drdobbs.com/architecture-and-design/the-bmp-file-format-part-1/184409517}
\showURL{%
\tempurl}


\bibitem[\protect\citeauthoryear{??}{wav}{2012}]%
        {wave_format}
 \bibinfo{year}{2012}\natexlab{}.
\newblock \bibinfo{title}{{WAVE Audio File Format}}.
\newblock
\newblock
\urldef\tempurl%
\url{https://www.loc.gov/preservation/digital/formats/fdd/fdd000001.shtml}
\showURL{%
\tempurl}


\bibitem[\protect\citeauthoryear{??}{chu}{2013}]%
        {chunked_cve}
 \bibinfo{year}{2013}\natexlab{}.
\newblock \bibinfo{title}{{CVE - CVE-2013-2028}}.
\newblock
\newblock
\urldef\tempurl%
\url{https://cve.mitre.org/cgi-bin/cvename.cgi?name=cve-2013-2028}
\showURL{%
\tempurl}


\bibitem[\protect\citeauthoryear{??}{bmp}{2018a}]%
        {bmp4_spec}
 \bibinfo{year}{2018}\natexlab{a}.
\newblock \bibinfo{title}{{BITMAPV4HEADER structure}}.
\newblock
\newblock
\urldef\tempurl%
\url{https://docs.microsoft.com/en-us/windows/win32/api/wingdi/ns-wingdi-bitmapv4header}
\showURL{%
\tempurl}


\bibitem[\protect\citeauthoryear{??}{bmp}{2018b}]%
        {bmp5_spec}
 \bibinfo{year}{2018}\natexlab{b}.
\newblock \bibinfo{title}{{BITMAPV5HEADER structure}}.
\newblock
\newblock
\urldef\tempurl%
\url{https://docs.microsoft.com/en-us/windows/win32/api/wingdi/ns-wingdi-bitmapv5header}
\showURL{%
\tempurl}


\bibitem[\protect\citeauthoryear{??}{goo}{2018}]%
        {google_yis_2018}
 \bibinfo{year}{2018}\natexlab{}.
\newblock \bibinfo{title}{{Google's Year in Search}}.
\newblock
\newblock
\urldef\tempurl%
\url{https://trends.google.com/trends/yis/2018/US/}
\showURL{%
\tempurl}


\bibitem[\protect\citeauthoryear{??}{rif}{2018}]%
        {riff}
 \bibinfo{year}{2018}\natexlab{}.
\newblock \bibinfo{title}{{RIFF}}.
\newblock
\newblock
\urldef\tempurl%
\url{http://fileformats.archiveteam.org/index.php?title=RIFF&oldid=36251}
\showURL{%
\tempurl}


\bibitem[\protect\citeauthoryear{Angluin}{Angluin}{1987}]%
        {angluin1987learning}
\bibfield{author}{\bibinfo{person}{Dana Angluin}.}
  \bibinfo{year}{1987}\natexlab{}.
\newblock \showarticletitle{Learning regular sets from queries and
  counterexamples}.
\newblock \bibinfo{journal}{\emph{Information and computation}}
  \bibinfo{volume}{75}, \bibinfo{number}{2} (\bibinfo{year}{1987}),
  \bibinfo{pages}{87--106}.
\newblock


\bibitem[\protect\citeauthoryear{Angluin}{Angluin}{1990}]%
        {angluin1990negative}
\bibfield{author}{\bibinfo{person}{Dana Angluin}.}
  \bibinfo{year}{1990}\natexlab{}.
\newblock \showarticletitle{Negative results for equivalence queries}.
\newblock \bibinfo{journal}{\emph{Machine Learning}} \bibinfo{volume}{5},
  \bibinfo{number}{2} (\bibinfo{year}{1990}), \bibinfo{pages}{121--150}.
\newblock


\bibitem[\protect\citeauthoryear{Bangert and Zeldovich}{Bangert and
  Zeldovich}{2014}]%
        {bangert2014nail}
\bibfield{author}{\bibinfo{person}{Julian Bangert} {and}
  \bibinfo{person}{Nickolai Zeldovich}.} \bibinfo{year}{2014}\natexlab{}.
\newblock \showarticletitle{Nail: A practical tool for parsing and generating
  data formats}. In \bibinfo{booktitle}{\emph{11th $\{$USENIX$\}$ Symposium on
  Operating Systems Design and Implementation ($\{$OSDI$\}$ 14)}}.
  \bibinfo{pages}{615--628}.
\newblock


\bibitem[\protect\citeauthoryear{Bastani, Sharma, Aiken, and Liang}{Bastani
  et~al\mbox{.}}{2017}]%
        {bastani2017synthesizing}
\bibfield{author}{\bibinfo{person}{Osbert Bastani}, \bibinfo{person}{Rahul
  Sharma}, \bibinfo{person}{Alex Aiken}, {and} \bibinfo{person}{Percy Liang}.}
  \bibinfo{year}{2017}\natexlab{}.
\newblock \showarticletitle{Synthesizing program input grammars}. In
  \bibinfo{booktitle}{\emph{ACM SIGPLAN Notices}}, Vol.~\bibinfo{volume}{52}.
  ACM, \bibinfo{pages}{95--110}.
\newblock


\bibitem[\protect\citeauthoryear{Bittau, Belay, Mashtizadeh, Mazi{\`e}res, and
  Boneh}{Bittau et~al\mbox{.}}{2014}]%
        {bittau2014hacking}
\bibfield{author}{\bibinfo{person}{Andrea Bittau}, \bibinfo{person}{Adam
  Belay}, \bibinfo{person}{Ali Mashtizadeh}, \bibinfo{person}{David
  Mazi{\`e}res}, {and} \bibinfo{person}{Dan Boneh}.}
  \bibinfo{year}{2014}\natexlab{}.
\newblock \showarticletitle{Hacking blind}. In \bibinfo{booktitle}{\emph{2014
  IEEE Symposium on Security and Privacy}}. IEEE, \bibinfo{pages}{227--242}.
\newblock


\bibitem[\protect\citeauthoryear{Blazytko, Bishop, Aschermann, Cappos,
  Schl{\"o}gel, Korshun, Abbasi, Schweighauser, Schinzel, Schumilo,
  et~al\mbox{.}}{Blazytko et~al\mbox{.}}{2019}]%
        {grimoire}
\bibfield{author}{\bibinfo{person}{Tim Blazytko}, \bibinfo{person}{Matt
  Bishop}, \bibinfo{person}{Cornelius Aschermann}, \bibinfo{person}{Justin
  Cappos}, \bibinfo{person}{Moritz Schl{\"o}gel}, \bibinfo{person}{Nadia
  Korshun}, \bibinfo{person}{Ali Abbasi}, \bibinfo{person}{Marco
  Schweighauser}, \bibinfo{person}{Sebastian Schinzel}, \bibinfo{person}{Sergej
  Schumilo}, {et~al\mbox{.}}} \bibinfo{year}{2019}\natexlab{}.
\newblock \showarticletitle{$\{$GRIMOIRE$\}$: Synthesizing structure while
  fuzzing}. In \bibinfo{booktitle}{\emph{28th $\{$USENIX$\}$ Security Symposium
  ($\{$USENIX$\}$ Security 19)}}. \bibinfo{pages}{1985--2002}.
\newblock


\bibitem[\protect\citeauthoryear{Caballero, Poosankam, Kreibich, and
  Song}{Caballero et~al\mbox{.}}{2009}]%
        {caballero2009dispatcher}
\bibfield{author}{\bibinfo{person}{Juan Caballero}, \bibinfo{person}{Pongsin
  Poosankam}, \bibinfo{person}{Christian Kreibich}, {and} \bibinfo{person}{Dawn
  Song}.} \bibinfo{year}{2009}\natexlab{}.
\newblock \showarticletitle{Dispatcher: Enabling active botnet infiltration
  using automatic protocol reverse-engineering}. In
  \bibinfo{booktitle}{\emph{Proceedings of the 16th ACM conference on Computer
  and communications security}}. ACM, \bibinfo{pages}{621--634}.
\newblock


\bibitem[\protect\citeauthoryear{Caballero and Song}{Caballero and
  Song}{2013}]%
        {caballero2013automatic}
\bibfield{author}{\bibinfo{person}{Juan Caballero} {and} \bibinfo{person}{Dawn
  Song}.} \bibinfo{year}{2013}\natexlab{}.
\newblock \showarticletitle{Automatic protocol reverse-engineering: Message
  format extraction and field semantics inference}.
\newblock \bibinfo{journal}{\emph{Computer Networks}} \bibinfo{volume}{57},
  \bibinfo{number}{2} (\bibinfo{year}{2013}), \bibinfo{pages}{451--474}.
\newblock


\bibitem[\protect\citeauthoryear{Caballero, Yin, Liang, and Song}{Caballero
  et~al\mbox{.}}{2007}]%
        {caballero2007polyglot}
\bibfield{author}{\bibinfo{person}{Juan Caballero}, \bibinfo{person}{Heng Yin},
  \bibinfo{person}{Zhenkai Liang}, {and} \bibinfo{person}{Dawn Song}.}
  \bibinfo{year}{2007}\natexlab{}.
\newblock \showarticletitle{Polyglot: Automatic extraction of protocol message
  format using dynamic binary analysis}. In
  \bibinfo{booktitle}{\emph{Proceedings of the 14th ACM conference on Computer
  and communications security}}. ACM, \bibinfo{pages}{317--329}.
\newblock


\bibitem[\protect\citeauthoryear{Cambronero, Dang, Vasilakis, Shen, Wu, and
  Rinard}{Cambronero et~al\mbox{.}}{2019}]%
        {alse}
\bibfield{author}{\bibinfo{person}{Jos\'{e}~P. Cambronero},
  \bibinfo{person}{Thurston~H.Y. Dang}, \bibinfo{person}{Nikos Vasilakis},
  \bibinfo{person}{Jiasi Shen}, \bibinfo{person}{Jerry Wu}, {and}
  \bibinfo{person}{Martin Rinard}.} \bibinfo{year}{2019}\natexlab{}.
\newblock \showarticletitle{{Active Learning for Software Engineering}}. In
  \bibinfo{booktitle}{\emph{SPLASH Onward!}}
\newblock


\bibitem[\protect\citeauthoryear{Comparetti, Wondracek, Kruegel, and
  Kirda}{Comparetti et~al\mbox{.}}{2009}]%
        {prospex}
\bibfield{author}{\bibinfo{person}{Paolo~Milani Comparetti},
  \bibinfo{person}{Gilbert Wondracek}, \bibinfo{person}{Christopher Kruegel},
  {and} \bibinfo{person}{Engin Kirda}.} \bibinfo{year}{2009}\natexlab{}.
\newblock \showarticletitle{{Prospex: Protocol specification extraction}}. In
  \bibinfo{booktitle}{\emph{2009 30th IEEE Symposium on Security and Privacy}}.
  IEEE, \bibinfo{pages}{110--125}.
\newblock


\bibitem[\protect\citeauthoryear{Cui, Kannan, and Wang}{Cui
  et~al\mbox{.}}{2007}]%
        {discoverer}
\bibfield{author}{\bibinfo{person}{Weidong Cui}, \bibinfo{person}{Jayanthkumar
  Kannan}, {and} \bibinfo{person}{Helen~J Wang}.}
  \bibinfo{year}{2007}\natexlab{}.
\newblock \showarticletitle{{Discoverer: Automatic Protocol Reverse Engineering
  from Network Traces.}}. In \bibinfo{booktitle}{\emph{USENIX Security
  Symposium}}. \bibinfo{pages}{1--14}.
\newblock


\bibitem[\protect\citeauthoryear{Cui, Peinado, Chen, Wang, and Irun-Briz}{Cui
  et~al\mbox{.}}{2008}]%
        {tupni}
\bibfield{author}{\bibinfo{person}{Weidong Cui}, \bibinfo{person}{Marcus
  Peinado}, \bibinfo{person}{Karl Chen}, \bibinfo{person}{Helen~J Wang}, {and}
  \bibinfo{person}{Luis Irun-Briz}.} \bibinfo{year}{2008}\natexlab{}.
\newblock \showarticletitle{{Tupni: Automatic reverse engineering of input
  formats}}. In \bibinfo{booktitle}{\emph{Proceedings of the 15th ACM
  conference on Computer and communications security}}.
  \bibinfo{pages}{391--402}.
\newblock


\bibitem[\protect\citeauthoryear{De~la Higuera}{De~la Higuera}{2010}]%
        {de2010grammatical}
\bibfield{author}{\bibinfo{person}{Colin De~la Higuera}.}
  \bibinfo{year}{2010}\natexlab{}.
\newblock \bibinfo{booktitle}{\emph{Grammatical inference: learning automata
  and grammars}}.
\newblock \bibinfo{publisher}{Cambridge University Press}.
\newblock


\bibitem[\protect\citeauthoryear{De~Moura and Bj{\o}rner}{De~Moura and
  Bj{\o}rner}{2008}]%
        {z3}
\bibfield{author}{\bibinfo{person}{Leonardo De~Moura} {and}
  \bibinfo{person}{Nikolaj Bj{\o}rner}.} \bibinfo{year}{2008}\natexlab{}.
\newblock \showarticletitle{Z3: An efficient SMT solver}. In
  \bibinfo{booktitle}{\emph{International conference on Tools and Algorithms
  for the Construction and Analysis of Systems}}. Springer,
  \bibinfo{pages}{337--340}.
\newblock


\bibitem[\protect\citeauthoryear{Endignoux, Levillain, and Migeon}{Endignoux
  et~al\mbox{.}}{2016}]%
        {endignoux2016caradoc}
\bibfield{author}{\bibinfo{person}{Guillaume Endignoux},
  \bibinfo{person}{Olivier Levillain}, {and} \bibinfo{person}{Jean-Yves
  Migeon}.} \bibinfo{year}{2016}\natexlab{}.
\newblock \showarticletitle{Caradoc: a pragmatic approach to PDF parsing and
  validation}. In \bibinfo{booktitle}{\emph{2016 IEEE Security and Privacy
  Workshops (SPW)}}. Ieee, \bibinfo{pages}{126--139}.
\newblock


\bibitem[\protect\citeauthoryear{{Fars Robotics Website}}{{Fars Robotics
  Website}}{nd}]%
        {mt7610_drivers}
\bibfield{author}{\bibinfo{person}{{Fars Robotics Website}}.}
  \bibinfo{year}{n.d.}\natexlab{}.
\newblock \bibinfo{title}{{mt7610-drivers}}.
\newblock
\newblock
\urldef\tempurl%
\url{http://downloads.fars-robotics.net/wifi-drivers/mt7610-drivers/}
\showURL{%
\tempurl}


\bibitem[\protect\citeauthoryear{Gold}{Gold}{1978}]%
        {gold1978complexity}
\bibfield{author}{\bibinfo{person}{E~Mark Gold}.}
  \bibinfo{year}{1978}\natexlab{}.
\newblock \showarticletitle{Complexity of automaton identification from given
  data}.
\newblock \bibinfo{journal}{\emph{Information and control}}
  \bibinfo{volume}{37}, \bibinfo{number}{3} (\bibinfo{year}{1978}),
  \bibinfo{pages}{302--320}.
\newblock


\bibitem[\protect\citeauthoryear{Gopinath, Mathis, and Zeller}{Gopinath
  et~al\mbox{.}}{2020}]%
        {mining_input_grammars}
\bibfield{author}{\bibinfo{person}{Rahul Gopinath}, \bibinfo{person}{Bj{\"o}rn
  Mathis}, {and} \bibinfo{person}{Andreas Zeller}.}
  \bibinfo{year}{2020}\natexlab{}.
\newblock \showarticletitle{{Mining Input Grammars from Dynamic Control Flow}}.
  In \bibinfo{booktitle}{\emph{Proceedings of the ACM Joint European Software
  Engineering Conference and Symposium on the Foundations of Software
  Engineering (ESEC/FSE)}}.
\newblock


\bibitem[\protect\citeauthoryear{H{\"o}schele and Zeller}{H{\"o}schele and
  Zeller}{2016}]%
        {hoschele2016mining}
\bibfield{author}{\bibinfo{person}{Matthias H{\"o}schele} {and}
  \bibinfo{person}{Andreas Zeller}.} \bibinfo{year}{2016}\natexlab{}.
\newblock \showarticletitle{Mining input grammars from dynamic taints}. In
  \bibinfo{booktitle}{\emph{Proceedings of the 31st IEEE/ACM International
  Conference on Automated Software Engineering}}. ACM,
  \bibinfo{pages}{720--725}.
\newblock


\bibitem[\protect\citeauthoryear{James, Witten, Hastie, and Tibshirani}{James
  et~al\mbox{.}}{2013}]%
        {isl}
\bibfield{author}{\bibinfo{person}{Gareth James}, \bibinfo{person}{Daniela
  Witten}, \bibinfo{person}{Trevor Hastie}, {and} \bibinfo{person}{Robert
  Tibshirani}.} \bibinfo{year}{2013}\natexlab{}.
\newblock \bibinfo{booktitle}{\emph{An introduction to statistical learning}}.
  Vol.~\bibinfo{volume}{112}.
\newblock \bibinfo{publisher}{Springer}.
\newblock


\bibitem[\protect\citeauthoryear{Long, Ganesh, Carbin, Sidiroglou, and
  Rinard}{Long et~al\mbox{.}}{2012}]%
        {InputRectification}
\bibfield{author}{\bibinfo{person}{Fan Long}, \bibinfo{person}{Vijay Ganesh},
  \bibinfo{person}{Michael Carbin}, \bibinfo{person}{Stelios Sidiroglou}, {and}
  \bibinfo{person}{Martin~C. Rinard}.} \bibinfo{year}{2012}\natexlab{}.
\newblock \showarticletitle{Automatic input rectification}. In
  \bibinfo{booktitle}{\emph{34th International Conference on Software
  Engineering, {ICSE} 2012, June 2-9, 2012, Zurich, Switzerland}}.
  \bibinfo{pages}{80--90}.
\newblock


\bibitem[\protect\citeauthoryear{Long, Sidiroglou{-}Douskos, and Rinard}{Long
  et~al\mbox{.}}{2014}]%
        {RecoveryShepherding}
\bibfield{author}{\bibinfo{person}{Fan Long}, \bibinfo{person}{Stelios
  Sidiroglou{-}Douskos}, {and} \bibinfo{person}{Martin~C. Rinard}.}
  \bibinfo{year}{2014}\natexlab{}.
\newblock \showarticletitle{Automatic runtime error repair and containment via
  recovery shepherding}. In \bibinfo{booktitle}{\emph{{ACM} {SIGPLAN}
  Conference on Programming Language Design and Implementation, {PLDI} '14,
  Edinburgh, United Kingdom - June 09 - 11, 2014}}. \bibinfo{pages}{227--238}.
\newblock


\bibitem[\protect\citeauthoryear{Mendis, Bosboom, Wu, Kamil, Ragan-Kelley,
  Paris, Zhao, and Amarasinghe}{Mendis et~al\mbox{.}}{2015}]%
        {mendis2015helium}
\bibfield{author}{\bibinfo{person}{Charith Mendis}, \bibinfo{person}{Jeffrey
  Bosboom}, \bibinfo{person}{Kevin Wu}, \bibinfo{person}{Shoaib Kamil},
  \bibinfo{person}{Jonathan Ragan-Kelley}, \bibinfo{person}{Sylvain Paris},
  \bibinfo{person}{Qin Zhao}, {and} \bibinfo{person}{Saman Amarasinghe}.}
  \bibinfo{year}{2015}\natexlab{}.
\newblock \showarticletitle{{Helium: lifting high-performance stencil kernels
  from stripped x86 binaries to halide DSL code}}. In
  \bibinfo{booktitle}{\emph{Proceedings of the 36th ACM SIGPLAN Conference on
  Programming Language Design and Implementation}}. \bibinfo{pages}{391--402}.
\newblock


\bibitem[\protect\citeauthoryear{{Microsoft Research}}{{Microsoft
  Research}}{[n.d.]}]%
        {z3tutorial}
\bibfield{author}{\bibinfo{person}{{Microsoft Research}}.}
  \bibinfo{year}{[n.d.]}\natexlab{}.
\newblock \bibinfo{title}{{Getting Started with Z3: A Guide}}.
\newblock
\newblock
\urldef\tempurl%
\url{https://rise4fun.com/z3/tutorial#h21}
\showURL{%
\tempurl}


\bibitem[\protect\citeauthoryear{Momot, Bratus, Hallberg, and Patterson}{Momot
  et~al\mbox{.}}{2016}]%
        {momot2016seven}
\bibfield{author}{\bibinfo{person}{Falcon Momot}, \bibinfo{person}{Sergey
  Bratus}, \bibinfo{person}{Sven~M Hallberg}, {and} \bibinfo{person}{Meredith~L
  Patterson}.} \bibinfo{year}{2016}\natexlab{}.
\newblock \showarticletitle{The seven turrets of babel: A taxonomy of langsec
  errors and how to expunge them}. In \bibinfo{booktitle}{\emph{2016 IEEE
  Cybersecurity Development (SecDev)}}. IEEE, \bibinfo{pages}{45--52}.
\newblock


\bibitem[\protect\citeauthoryear{OpenWrt}{OpenWrt}{2021}]%
        {openwrt}
\bibfield{author}{\bibinfo{person}{OpenWrt}.} \bibinfo{year}{2021}\natexlab{}.
\newblock \bibinfo{title}{{mac80211 driver for MediaTek MT76x0e, MT76x2e,
  MT7603, MT7615, MT7628 and MT768}}.
\newblock
\newblock
\urldef\tempurl%
\url{https://github.com/openwrt/mt76}
\showURL{%
\tempurl}


\bibitem[\protect\citeauthoryear{{Perl 5 Porters}}{{Perl 5 Porters}}{[n.d.]a}]%
        {perlnumber}
\bibfield{author}{\bibinfo{person}{{Perl 5 Porters}}.}
  \bibinfo{year}{[n.d.]}\natexlab{a}.
\newblock \bibinfo{title}{{perlnumber - semantics of numbers and numeric
  operations in Perl}}.
\newblock
\newblock
\urldef\tempurl%
\url{https://perldoc.perl.org/perlnumber}
\showURL{%
\tempurl}


\bibitem[\protect\citeauthoryear{{Perl 5 Porters}}{{Perl 5 Porters}}{[n.d.]b}]%
        {perlop}
\bibfield{author}{\bibinfo{person}{{Perl 5 Porters}}.}
  \bibinfo{year}{[n.d.]}\natexlab{b}.
\newblock \bibinfo{title}{{perlop - Perl Operators and Precedence}}.
\newblock
\newblock
\urldef\tempurl%
\url{https://perldoc.perl.org/perlop#Shift-Operators}
\showURL{%
\tempurl}


\bibitem[\protect\citeauthoryear{Polozov and Gulwani}{Polozov and
  Gulwani}{2015}]%
        {polozov2015flashmeta}
\bibfield{author}{\bibinfo{person}{Oleksandr Polozov} {and}
  \bibinfo{person}{Sumit Gulwani}.} \bibinfo{year}{2015}\natexlab{}.
\newblock \showarticletitle{FlashMeta: a framework for inductive program
  synthesis}. In \bibinfo{booktitle}{\emph{ACM SIGPLAN Notices}},
  Vol.~\bibinfo{volume}{50}. ACM, \bibinfo{pages}{107--126}.
\newblock


\bibitem[\protect\citeauthoryear{Raza and Gulwani}{Raza and Gulwani}{2018}]%
        {raza2018disjunctive}
\bibfield{author}{\bibinfo{person}{Mohammad Raza} {and} \bibinfo{person}{Sumit
  Gulwani}.} \bibinfo{year}{2018}\natexlab{}.
\newblock \showarticletitle{Disjunctive Program Synthesis: A Robust Approach to
  Programming by Example}. In \bibinfo{booktitle}{\emph{Thirty-Second AAAI
  Conference on Artificial Intelligence}}.
\newblock


\bibitem[\protect\citeauthoryear{Sapp}{Sapp}{nd}]%
        {wave_pcm}
\bibfield{author}{\bibinfo{person}{Craig~Stuart Sapp}.}
  \bibinfo{year}{n.d.}\natexlab{}.
\newblock \bibinfo{title}{{WAVE PCM soundfile format}}.
\newblock
\newblock
\urldef\tempurl%
\url{http://soundfile.sapp.org/doc/WaveFormat/}
\showURL{%
\tempurl}


\bibitem[\protect\citeauthoryear{Serebryany, Bruening, Potapenko, and
  Vyukov}{Serebryany et~al\mbox{.}}{2012}]%
        {address_sanitizer}
\bibfield{author}{\bibinfo{person}{Konstantin Serebryany},
  \bibinfo{person}{Derek Bruening}, \bibinfo{person}{Alexander Potapenko},
  {and} \bibinfo{person}{Dmitriy Vyukov}.} \bibinfo{year}{2012}\natexlab{}.
\newblock \showarticletitle{{AddressSanitizer: A fast address sanity checker}}.
  In \bibinfo{booktitle}{\emph{Presented as part of the 2012 $\{$USENIX$\}$
  Annual Technical Conference ($\{$USENIX$\}$$\{$ATC$\}$ 12)}}.
  \bibinfo{pages}{309--318}.
\newblock


\bibitem[\protect\citeauthoryear{Settles}{Settles}{2009}]%
        {settles2009active}
\bibfield{author}{\bibinfo{person}{Burr Settles}.}
  \bibinfo{year}{2009}\natexlab{}.
\newblock \bibinfo{booktitle}{\emph{Active learning literature survey}}.
\newblock \bibinfo{type}{{T}echnical {R}eport}.
  \bibinfo{institution}{University of Wisconsin-Madison Department of Computer
  Sciences}.
\newblock


\bibitem[\protect\citeauthoryear{Shen and Rinard}{Shen and Rinard}{2017}]%
        {FilteredIterator}
\bibfield{author}{\bibinfo{person}{Jiasi Shen} {and} \bibinfo{person}{Martin
  Rinard}.} \bibinfo{year}{2017}\natexlab{}.
\newblock \showarticletitle{Robust Programs with Filtered Iterators}. In
  \bibinfo{booktitle}{\emph{Proceedings of the 10th ACM SIGPLAN International
  Conference on Software Language Engineering}} (Vancouver, BC, Canada)
  \emph{(\bibinfo{series}{SLE 2017})}. \bibinfo{pages}{244--255}.
\newblock
\showISBNx{978-1-4503-5525-4}


\bibitem[\protect\citeauthoryear{Shen and Rinard}{Shen and Rinard}{2019}]%
        {konure}
\bibfield{author}{\bibinfo{person}{Jiasi Shen} {and} \bibinfo{person}{Martin~C
  Rinard}.} \bibinfo{year}{2019}\natexlab{}.
\newblock \showarticletitle{Using active learning to synthesize models of
  applications that access databases}. In \bibinfo{booktitle}{\emph{Proceedings
  of the 40th ACM SIGPLAN Conference on Programming Language Design and
  Implementation}}. ACM, \bibinfo{pages}{269--285}.
\newblock


\bibitem[\protect\citeauthoryear{Sidiroglou{-}Douskos, Lahtinen, Long, and
  Rinard}{Sidiroglou{-}Douskos et~al\mbox{.}}{2015}]%
        {CodePhage}
\bibfield{author}{\bibinfo{person}{Stelios Sidiroglou{-}Douskos},
  \bibinfo{person}{Eric Lahtinen}, \bibinfo{person}{Fan Long}, {and}
  \bibinfo{person}{Martin Rinard}.} \bibinfo{year}{2015}\natexlab{}.
\newblock \showarticletitle{Automatic error elimination by horizontal code
  transfer across multiple applications}. In
  \bibinfo{booktitle}{\emph{Proceedings of the 36th {ACM} {SIGPLAN} Conference
  on Programming Language Design and Implementation, Portland, OR, USA, June
  15-17, 2015}}. \bibinfo{pages}{43--54}.
\newblock


\bibitem[\protect\citeauthoryear{Sidiroglou-Douskos, Lahtinen, Rittenhouse,
  Piselli, Long, Kim, and Rinard}{Sidiroglou-Douskos et~al\mbox{.}}{2015}]%
        {diode}
\bibfield{author}{\bibinfo{person}{Stelios Sidiroglou-Douskos},
  \bibinfo{person}{Eric Lahtinen}, \bibinfo{person}{Nathan Rittenhouse},
  \bibinfo{person}{Paolo Piselli}, \bibinfo{person}{Fan Long},
  \bibinfo{person}{Deokhwan Kim}, {and} \bibinfo{person}{Martin Rinard}.}
  \bibinfo{year}{2015}\natexlab{}.
\newblock \showarticletitle{Targeted automatic integer overflow discovery using
  goal-directed conditional branch enforcement}. In
  \bibinfo{booktitle}{\emph{ACM Sigplan Notices}}, Vol.~\bibinfo{volume}{50}.
  ACM, \bibinfo{pages}{473--486}.
\newblock


\bibitem[\protect\citeauthoryear{Singh}{Singh}{2016}]%
        {singh2016blinkfill}
\bibfield{author}{\bibinfo{person}{Rishabh Singh}.}
  \bibinfo{year}{2016}\natexlab{}.
\newblock \showarticletitle{Blinkfill: Semi-supervised programming by example
  for syntactic string transformations}.
\newblock \bibinfo{journal}{\emph{Proceedings of the VLDB Endowment}}
  \bibinfo{volume}{9}, \bibinfo{number}{10} (\bibinfo{year}{2016}),
  \bibinfo{pages}{816--827}.
\newblock


\bibitem[\protect\citeauthoryear{Underwood and Locasto}{Underwood and
  Locasto}{2016}]%
        {underwood2016search}
\bibfield{author}{\bibinfo{person}{Katherine Underwood} {and}
  \bibinfo{person}{Michael~E Locasto}.} \bibinfo{year}{2016}\natexlab{}.
\newblock \showarticletitle{In Search of Shotgun Parsers in Android
  Applications}. In \bibinfo{booktitle}{\emph{2016 IEEE Security and Privacy
  Workshops (SPW)}}. IEEE, \bibinfo{pages}{140--155}.
\newblock


\bibitem[\protect\citeauthoryear{Vasa}{Vasa}{2019}]%
        {google_images_download}
\bibfield{author}{\bibinfo{person}{Hardik Vasa}.}
  \bibinfo{year}{2019}\natexlab{}.
\newblock \bibinfo{title}{{google\_images\_download}}.
\newblock
\newblock
\urldef\tempurl%
\url{https://github.com/hardikvasa/google-images-download}
\showURL{%
\tempurl}


\bibitem[\protect\citeauthoryear{Wang, Zhang, Yao, Qu, and Guo}{Wang
  et~al\mbox{.}}{2011}]%
        {veritas}
\bibfield{author}{\bibinfo{person}{Yipeng Wang}, \bibinfo{person}{Zhibin
  Zhang}, \bibinfo{person}{Danfeng~Daphne Yao}, \bibinfo{person}{Buyun Qu},
  {and} \bibinfo{person}{Li Guo}.} \bibinfo{year}{2011}\natexlab{}.
\newblock \showarticletitle{{Inferring protocol state machine from network
  traces: a probabilistic approach}}. In
  \bibinfo{booktitle}{\emph{International Conference on Applied Cryptography
  and Network Security}}. Springer, \bibinfo{pages}{1--18}.
\newblock


\bibitem[\protect\citeauthoryear{Wu}{Wu}{2018}]%
        {wu}
\bibfield{author}{\bibinfo{person}{Jerry Wu}.} \bibinfo{year}{2018}\natexlab{}.
\newblock \bibinfo{booktitle}{\emph{Using Dynamic Analysis to Infer Python
  Programs and Convert Them into Database Programs.}}
\newblock \bibinfo{publisher}{MIT}.
\newblock


\end{thebibliography}

\newpage

\begin{appendices}

\section{Appendices}
\subsection{\diode Expression Denouement}
\label{sec:example_diode}

We walk through part of the \diode expression for extracting a green subpixel
from a 16-bit bitmap.

\noindent\textbf{Background: 16-bit bitmaps.} 16-bit bitmaps bitpack the red, green, and blue subpixels into
two bytes, with 5-bits for each of the colors\footnote{\slime can also handle variants such as 5-6-5}:

\begin{lstlisting}[basicstyle=\blobformat, escapeinside={<@}{@>}]
    -<@\textcolor{red}{RRRRR}@><@\textcolor{green}{GG}@> <@\textcolor{green}{GGG}@><@\textcolor{blue}{BBBBB}@>
\end{lstlisting}

Notably, the green subpixel spans both bytes. The goal is to convert the green subpixel into an 8-bit value (ranging from 0 to 255).

\noindent\textbf{\diode expression, part 1}. The first step is to read two adjacent bytes from the input file, and combine them into a 16-bit value. This is performed by the following portion of the \diode expression:

\begin{lstlisting}[basicstyle=\blobformat, escapeinside={<@}{@>}]
(let ((?x3567
   (bvor
      (bvshl
         ((_ zero_extend 24)
            value_0_0x189_0x189)
         (_ bv8 32))
      ((_ zero_extend 24)
         value_0_0x188_0x188))))
\end{lstlisting}

The first byte (\lstinline[basicstyle=\ttfamily, escapeinside={<@}{@>}]!value_0_0x188_0x188_little!; offsets refer to bytes in
the input file) contains:
\begin{lstlisting}[basicstyle=\blobformat, escapeinside={<@}{@>}]
    -<@\textcolor{red}{RRRRR}@><@\textcolor{green}{GG}@>
\end{lstlisting}
\noindent and the second byte (\lstinline[basicstyle=\ttfamily, escapeinside={<@}{@>}]!value_0_0x189_0x189_little!) contains
\begin{lstlisting}[basicstyle=\blobformat, escapeinside={<@}{@>}]
    <@\textcolor{green}{GGG}@><@\textcolor{blue}{BBBBB}@>
\end{lstlisting}

The \code{let} expression concatenates these bytes into a 16-bit word,
using a bitvector shift left (``bvshl'') by 8-bits (``bv8'') of the
second-byte, followed by a bitwise OR (``bvor'') with the first-byte.
The byte order is reversed due to the little-endian x86 architecture.

The value \code{?x3567} is internally represented as 32-bit\footnote{For brevity, we will present them below as 16-bit.} (to prevent overflow in later steps), which necessitates the explicit zero-extensions because \smtlib is strongly typed:

\begin{lstlisting}[basicstyle=\blobformat, escapeinside={<@}{@>}]
    -<@\textcolor{red}{RRRRR}@><@\textcolor{green}{GGGGG}@><@\textcolor{blue}{BBBBB}@>
\end{lstlisting}

We now continue with how to extract the green subpixel as an 8-bit value.

\noindent\textbf{\diode expression, part 2}

\begin{lstlisting}[basicstyle=\blobformat, escapeinside={<@}{@>}]
  (let
    (
       (?x3612
         (bvashr
           (bvmul
             (_ bv33 32)
             (bvashr
              (bvashr
                (bvand ?x3567 (_ bv992 32))
                (_ bv2 32))
              (_ bv3 32)))
           (_ bv2 32))))
\end{lstlisting}

The innermost portion of the expression:
\begin{lstlisting}[basicstyle=\blobformat, escapeinside={<@}{@>}]
    (bvand ?x3567 (_ bv992 32))
\end{lstlisting}
extracts the green subpixel, by performing a bit-wise AND with the
bit-vector of value 992 (\lstinline[basicstyle=\ttfamily, escapeinside={<@}{@>}]!bv992!; \lstinline[basicstyle=\ttfamily, escapeinside={<@}{@>}]!0000001111100000! in binary); thus, the
expression is equal to:
\begin{lstlisting}[basicstyle=\blobformat, escapeinside={<@}{@>}]
    000000<@\textcolor{green}{GGGGG}@>00000
\end{lstlisting}

The bitvector arithmetic shift right (\lstinline[basicstyle=\ttfamily, escapeinside={<@}{@>}]!bvashr!) by 5-bits (in two
separate shifts of 3- and 2-bits due to arcane compiler optimizations)
results in the bitvector:
\begin{lstlisting}[basicstyle=\blobformat, escapeinside={<@}{@>}]
    00000000000<@\textcolor{green}{GGGGG}@>
\end{lstlisting}
\noindent Notice that the sign-extension
arising from the arithmetic (as opposed to logical) shift right is
equivalent to zero-extension in this case, since the leftmost bit was zero.

If we directly used the green value, it would range from 0 to 31.
A \naive approach of bit-shifting left by 3-bits would result in a
range of 0 to 248, which is less than the full range of an 8-bit value
(0 to 255). A less \naive approach would therefore multiply by 255
and divide by 31, but that is inefficient. Instead, \stbimage multiplies
by 33, then bit-shifts right by 2-bits, taking advantage of the identity:
\begin{lstlisting}[basicstyle=\blobformat, escapeinside={<@}{@>}]
    ((unsigned) (31 * 33) >> 2) == 255
\end{lstlisting}
This results in:
\begin{lstlisting}[basicstyle=\blobformat, escapeinside={<@}{@>}]
    00000000<@\textcolor{green}{GGGGGGGG}@>
\end{lstlisting}

\noindent\textbf{\diode expression, part 3}

\begin{lstlisting}[basicstyle=\blobformat, escapeinside={<@}{@>}]
  (let
    (
      (?x481039
        (
          (_ sign_extend 56)
          ((_ extract 7 0) ?x3612))))))))
\end{lstlisting}

The third let expression extracts the eight right-most bits
(from indices 0 to 7; note that the high index is listed before the low
index), then sign extends it to be a 64-bit value.

\subsection{Excerpt of \diode log}
\label{sec:diode_log}

Below is an excerpt of a \diode log file from running \catimg on a 16-bit bottom-up bitmap, after pre-processing into JSON by their \code{dfsan\_parse} utility. Notice that it is a byte-by-byte dump of the output data structure, using hardcoded input file indices, and lacks any direct information about the looping structure of the input file or output data structure. For example, it is not immediately obvious if any expressions are identical (after abstracting away the concrete input indices denoted by \code{**value\_...}), nor that the output bytes belong to the same data structure (\diode can instrument multiple variables/data structures, which are stored in a combined log). For an explanation of the \code{expression} field, consult Section \ref{sec:example_diode} or your local Z3 dealer~\cite{z3}.

\begin{lstlisting}[basicstyle=\blobformat, escapeinside={<@}{@>}]
...
{
     "address": 94800882445552,
     "constraints": [],
     "expression":
         "; \n
          (set-info :status unknown)\n
          (declare-fun **value_0_0x43e_0x43e_little () (_ BitVec 8))\n
          (declare-fun **value_0_0x43f_0x43f_little () (_ BitVec 8))\n
          (declare-fun __dfsan_top_level () (_ BitVec 64))\n
          (assert\n
          (let ((?x226 (bvsub (_ bv0 32) (_ bv4294967293 32))))\n
          (let ((?x13030
               (bvor (bvshl
                   ((_ zero_extend 24) **value_0_0x43f_0x43f_little) (_ bv8 32))
                   ((_ zero_extend 24) **value_0_0x43e_0x43e_little))))\n
          (let ((?x13047
               (bvashr (bvmul (_ bv33 32) (bvashr (bvshl (bvand ?x13030
               (_ bv31 32)) ?x226) (_ bv3 32))) (_ bv2 32))))\n
          (let ((?x4865826 (bvand ((_ sign_extend 56)
               ((_ extract 7 0) ?x13047)) (_ bv65535 64))))\n
          (= __dfsan_top_level ?x4865826))))))\n
          (check-sat)\n",
    "id": 1299951925357182983,
    "sink_type": "malloc",
    "size": 0,
    "stacktrace":
        [
             "catimg;dfsan_custom.cc@1101:__dfsw_malloc+238595",
             "catimg;useData.h@29:dfs$use_data0+263016",
             "catimg;useData.h@55:dfs$use_data_loop_logged+263861",
             "catimg;sh_image.c@123:dfs$stbi_xload+527240",
             "catimg;sh_image.c@219:dfs$img_load_from_file+534550",
             "catimg;catimg.c@115:main+251177",
             "libc.so.6;libc-start.c@291:__libc_start_main+133167",
             "catimg;<invalid>@0:_start+102440"
        ]
},

{
     "address": 94800882445552,
     "constraints": [],
     "expression":
         "; \n
          (set-info :status unknown)\n
          (declare-fun **value_0_0x440_0x440_little () (_ BitVec 8))\n
          (declare-fun **value_0_0x441_0x441_little () (_ BitVec 8))\n
          (declare-fun __dfsan_top_level () (_ BitVec 64))\n
          (assert\n
          (let ((?x13055 (bvor (bvshl ((_ zero_extend 24) 
               **value_0_0x441_0x441_little) (_ bv8 32))
               ((_ zero_extend 24) **value_0_0x440_0x440_little))))\n
          (let ((?x13060 (bvashr (bvmul (bvashr (bvashr
              (bvand ?x13055 (_ bv31744 32)) (_ bv7 32))
              (_ bv3 32)) (_ bv33 32)) (_ bv2 32))))\n
          (let ((?x4865828 (bvand ((_ sign_extend 56)
              ((_ extract 7 0) ?x13060)) (_ bv65535 64))))\n
          (= __dfsan_top_level ?x4865828)))))\n
          (check-sat)\n",
     "id": 1299951925357182983,
     "sink_type": "malloc",
     "size": 0,
     "stacktrace":
         [
             "catimg;dfsan_custom.cc@1101:__dfsw_malloc+238595",
             "catimg;useData.h@29:dfs$use_data0+263016",
             "catimg;useData.h@55:dfs$use_data_loop_logged+263861",
             "catimg;sh_image.c@123:dfs$stbi_xload+527240",
             "catimg;sh_image.c@219:dfs$img_load_from_file+534550",
             "catimg;catimg.c@115:main+251177",
             "libc.so.6;libc-start.c@291:__libc_start_main+133167",
             "catimg;<invalid>@0:_start+102440"
         ]
},

{
     "address": 94800882445552,
     "constraints": [],
     "expression":
         "; \n
          (set-info :status unknown)\n
          (declare-fun **value_0_0x440_0x440_little () (_ BitVec 8))\n
          (declare-fun **value_0_0x441_0x441_little () (_ BitVec 8))\n
          (declare-fun __dfsan_top_level () (_ BitVec 64))\n
          (assert\n
          (let ((?x13055 (bvor (bvshl ((_ zero_extend 24)
              **value_0_0x441_0x441_little) (_ bv8 32))
              ((_ zero_extend 24) **value_0_0x440_0x440_little))))\n
          (let ((?x13066 (bvashr (bvmul (_ bv33 32)
              (bvashr (bvashr (bvand ?x13055 (_ bv992 32)) (_ bv2 32))
              (_ bv3 32))) (_ bv2 32))))\n
          (let ((?x4865830 (bvand ((_ sign_extend 56)
              ((_ extract 7 0) ?x13066)) (_ bv65535 64))))\n
          (= __dfsan_top_level ?x4865830)))))\n
          (check-sat)\n",
     "id": 1299951925357182983,
     "sink_type": "malloc",
     "size": 0,
     "stacktrace":
         [
              "catimg;dfsan_custom.cc@1101:__dfsw_malloc+238595",
              "catimg;useData.h@29:dfs$use_data0+263016",
              "catimg;useData.h@55:dfs$use_data_loop_logged+263861",
              "catimg;sh_image.c@123:dfs$stbi_xload+527240",
              "catimg;sh_image.c@219:dfs$img_load_from_file+534550",
              "catimg;catimg.c@115:main+251177",
              "libc.so.6;libc-start.c@291:__libc_start_main+133167",
              "catimg;<invalid>@0:_start+102440"
         ]
},
\end{lstlisting}

\subsection{Wrapper Function to Read Input (for C Backend)}
\label{sec:wrapper_read_input}

This function assumes that, once we start calling this function,
there will be no intervening \code{fread} or \code{fseek} operations
(this allows avoiding the overhead of calling \code{ftell} and \code{fseek}).
This assumption true for \slime's regenerated parsers, which only perform
file reads via this wrapper function.

\begin{lstlisting}[basicstyle=\blobformat, escapeinside={<@}{@>}]
void readBytesFromFP (char* buf, FILE* fp, int start, int size) {
    static long oldPos = -1;
    if (oldPos == -1) {
        oldPos = ftell (fp);
    }
    if (oldPos == -1) {
        printf ("Error: oldPos == -1\n");
        abort ();
    }

    if (oldPos != start) {
        int sought;
        sought = fseek (fp, start, SEEK_SET);
        if (sought != 0) {
            printf ("Unable to fseek to %
                    start, sought);
            abort ();
        }

        oldPos = start;
    }

    ssize_t bytesRead = fread (buf, 1, size, fp);
    if (bytesRead != size) {
        printf ("Read %
        abort ();
    }

    oldPos = oldPos + bytesRead;
}
\end{lstlisting}

\subsection{Dynamic Array Wrapper Function (for C Backend)}
\label{sec:dynamic_array}

\begin{lstlisting}[basicstyle=\blobformat, escapeinside={<@}{@>}]
// writeArray : Stores the character into the array at the specified
//              index, resizing the array if necessary.
inline void writeArray (unsigned char** out, long* size, long index,
                        char newValue) {
    if (index < 0) {
        printf ("Invalid index %
        abort ();
    }

    if (index >= *size) {
        if (*size == 0) {
            *out = NULL; // Not guaranteed that it was initialized
        }

        long newSize = *size;

        while (index >= newSize) {
            // Should check when newSize is close to LONG_MAX
            // but on x64 we'll run out of memory before that happens
            newSize = (newSize + 1) * 2;
        }

        *size = newSize;
        assert (size > 0); // Catch wrap-around

        *out = realloc (*out, *size);
        if (*out == NULL) {
            printf ("Error: unable to realloc for %
                    *out, *size, index, newValue);
            abort ();
        }
    }

    (*out)[index] = newValue;
}
\end{lstlisting}

\subsection{Algorithms for Consistent Assignments}
\label{sec:consistent_assignments}

The steps for rewriting constants as expressions (Section \ref{sec:rewrite_constants}) and replacing remaining constants with header bytes (Section \ref{sec:replace_constants}) both require an algorithm for finding a set of assignments that are compatible with at least \emph{one} of the files. \slime's prototype uses a Cartesian product voting algorithm (described in Sectio \ref{sec:rewrite_constants}), which finds a set of assignments that is compatible with the most files. The Cartesian product algorithm has exponential runtime, but works acceptably in our case studies because of the small number of variables and allowed rewrites. A polynomial-time algorithm would allow \slime to scale to larger examples.\footnote{Note that if Cartesian product gets a ``better'' set of assignments, it might save overall CPU time by not needing \diode to process as many bytes of input.}

In this section, we present polynomial-time algorithms, some of which produce assignments which, in pilot experiments, were nearly as good as the Cartesian product voting algorithm. We compare the runtime complexity and assignment quality of these algorithms in Figure \ref{fig:assignment_complexities}. Note that we can potentially get tighter runtime bounds by counting the number of candidate assignments (the concatenation of ``variable = expression''), which is upper-bounded by $(\# variables) * (\# expressions)$.

\begin{figure}[!h]
\centering
\begin{tabular}{ | c | r |  r | }
\hline
\textbf{Algorithm} & \textbf{Runtime} & \textbf{Quality of Assignments} \\
\hline
Algorithm \ref{algo:consistent_assignment_simplest}: Simplest & $O(vef)$ & Meets Basic Requirement \\
\hline
Algorithm \ref{algo:consistent_assignment_conceptual}: Improved (conceptual) & $O((v^2)ef)$ & Better \\
Algorithm \ref{algo:consistent_assignment_optimized}: Improved (optimized) & $O((v^2)e + vef)$ & Better \\
\hline
Cartesian product voting & $O((e^v)f)$ & Best \\
\hline
\end{tabular}
\caption{
\textbf{
Comparison of consistent assignment algorithms' runtime complexity and quality of assignments.
}
}
\label{fig:assignment_complexities}
\end{figure}

\noindent\textbf{Simplest.} Randomly pick any one file, which will be designated the ``exemplar'' file. For each variable, randomly choose any expression that is compatible with the exemplar file. This is guaranteed to produce consistent assignments for at least that exemplar file. We formalize this in Algorithm \ref{algo:consistent_assignment_simplest}.

\begin{algorithm}
\caption{Finding a consistent set of files and variable/expression assignments (simplest)}
\label{algo:consistent_assignment_simplest}
\begin{algorithmic}
\Require{A list of files $\mathit{F}$, a list of variables $\mathit{V}$, a list of candidate expressions $\mathit{E}$ that can be assigned to the variables, a matrix of weights $\mathit{W}$ such that $W_{f,v,e}$ is zero if file $\mathit{f \in F}$ is incompatible with assigning expression $\mathit{e \in E}$ to variable $\mathit{v \in V}$ and positive otherwise. Files $\mathit{f \in F}$ do not need to contain all variables $\mathit{v \in V}$ \ie $\sum_{\mathit{e} \in \mathit{E}} W_{v,e,f}$ may equal zero.}

\Ensure{A non-empty list of $\mathit{compatibleFiles} \subseteq \mathit{F}$, and a list $\mathit{assignments}$ of 2-tuples $(\mathit{v} \in \mathit{V}, \mathit{e} \in \mathit{E})$ that fully specify the variables for each compatible file.}

\Function{ConsistentAssignment}{$\mathit{F}$, $\mathit{V}$, $\mathit{E}$, $\mathit{W}$}
    \State $\mathit{exemplarFile} \gets \mathit{F[0]}$ \Comment{Pick any file}

    \State $\mathit{assignments} \gets \emptyset$
    \State $\mathit{compatFiles} \gets \mathit{F}$
    
    \For{$\mathit{v} \in \mathit{V}$}
       \If{$\sum_{\mathit{e} \in \mathit{E}} W_{v,e,exemplarFile} > 0$}
            \State $e^{OPT} \gets \argmin_{\mathit{e} \in \mathit{E}} W_{v,e,exemplarFile} |  W_{v,e,exemplarFile} > 0$ \Comment{Any $e$ that works with $v$}
    
            \State $\mathit{assignments} \gets \mathit{assignments} + (\mathit{v}, \mathit{e^{OPT}})$
    
            \For{$\mathit{f} \in \mathit{compatFiles}$} \Comment{Remove any files that are incompatible with the assignment}
                \If{$\mathit{W_{v,e^{OPT},f}} = 0$}
                    \State $\mathit{compatFiles} \gets \mathit{compatFiles} \setminus \mathit{f}$
                \EndIf
            \EndFor
        \EndIf
    \EndFor

    \State \Return $(\mathit{compatFiles}, \mathit{assignments})$
\EndFunction
\end{algorithmic}
\end{algorithm}

\noindent\textbf{Improved (conceptual).}
Algorithm \ref{algo:consistent_assignment_conceptual} shows the conceptual algorithm for calculating consistent assignments. Initially, all files are compatible and all variables are unassigned. At each loop iteration, we select, from all expressions and the unassigned variables, the assignment (variable = expression) that has the most total weight across all compatible files. We then consider the variable assigned, and delete any files that are not compatible with this new assignment.

\begin{algorithm}
\caption{Finding a consistent set of files and variable/expression assignments (conceptual)}
\label{algo:consistent_assignment_conceptual}
\begin{algorithmic}
\Require{As per Algorithm \ref{algo:consistent_assignment_simplest}}

\Ensure{A non-empty list of $\mathit{compatibleFiles} \subseteq \mathit{F}$, and a list $\mathit{assignments}$ of 2-tuples $(\mathit{v} \in \mathit{V}, \mathit{e} \in \mathit{E})$ that fully specify the variables for each compatible file.}

\Function{ConsistentAssignment}{$\mathit{F}$, $\mathit{V}$, $\mathit{E}$, $\mathit{W}$}
    \State $\mathit{assignments} \gets \emptyset$
    \State $\mathit{compatFiles} \gets \mathit{F}$
    \State $\mathit{unassignedVars} \gets \mathit{V}$

    \While{$\sum_{f \in compatFiles, v \in \mathit{unassignedVars}, e \in E} W_{v,e,f} > 0$}
        \State $(\mathit{v^{OPT}}, \mathit{e^{OPT}}) \gets \argmax_{\mathit{v} \in \mathit{unassignedVars}, \mathit{e} \in \mathit{E}} \sum_{\mathit{f} \in \mathit{compatFiles}} W_{v,e,f}$
        \State $\mathit{assignments} \gets \mathit{assignments} + (\mathit{v^{OPT}}, \mathit{e^{OPT}})$

        \State $\mathit{unassignedVars} \gets \mathit{unassignedVars} \setminus \mathit{v^{OPT}}$
        
        \For{$\mathit{f} \in \mathit{compatFiles}$}
            \If{$\mathit{W_{v^{OPT},e^{OPT},f}} = 0$}
                \State $\mathit{compatFiles} \gets \mathit{compatFiles} \setminus \mathit{f}$
            \EndIf
        \EndFor
    \EndWhile
    
    \State \Return $(\mathit{compatFiles}, \mathit{assignments})$
\EndFunction
\end{algorithmic}
\end{algorithm}

\noindent\textbf{Improved (optimized).} There are two expensive steps in Algorithm \ref{algo:consistent_assignment_conceptual}, which scan across the entire 3D array:

$\sum_{f \in compatFiles, v \in \mathit{unassignedVars}, e \in E} W_{v,e,f}$

\noindent and

$\argmax_{\mathit{v} \in \mathit{unassignedVars}, \mathit{e} \in \mathit{E}} \sum_{\mathit{f} \in \mathit{compatFiles}} W_{v,e,f}.$

We observe that we can compute these quantities using a 2D array, where we have summed across the file dimension. This requires careful bookkeeping to update this 2D array whenever we remove a variable from the list of unassigned variables, or remove a file from the list of compatible files, but avoiding deleting the weight of a value twice (\ie once because the variable was assigned, and again if from an incompatible file). We present this in Algorithm \ref{algo:consistent_assignment_optimized}.

\begin{algorithm}
\caption{Finding a consistent set of files and variable/expression assignments (optimized)}
\label{algo:consistent_assignment_optimized}
\begin{algorithmic}
\Require{As per Algorithm \ref{algo:consistent_assignment_simplest}, except that we instantiate $\mathit{W}$ as nested hashtables.}
\Ensure{As per Algorithm \ref{algo:consistent_assignment_conceptual}}

\Function{Prune}{$\mathit{v}$, $\mathit{e}$, $\mathit{f}$, $\mathit{W}$, $\mathit{assignmentWeights}$}
    \State $\mathit{assignmentWeights}[v:e] \gets \mathit{assignmentWeights}[v:e] - \mathit{W}[v][e][f]$
    \If{$\mathit{assignmentWeights}[v:e] = 0$}
        \State delete $\mathit{assignmentWeights}[v:e]$
    \EndIf
\EndFunction

\Function{ConsistentAssignment}{$\mathit{F}$, $\mathit{V}$, $\mathit{E}$, $\mathit{W}$}
    \State $\mathit{compatibleFiles} \gets \emptyset$ \Comment{Place array in outer scope so that we can return it}
    \State $\mathit{assignmentWeights} \gets \emptyset, \mathit{assignmentsPerFile} \gets \emptyset$

    \For{$\mathit{v} \in \mathit{V}$} \Comment{Precompute summations of 3D-array}
        \For{$\mathit{e} \in ($keys $\mathit{W}[v])$}
            \For{$\mathit{f} \in ($keys $\mathit{W}[v][e])$}
                \State $\mathit{assignmentWeights}[v:e] \gets \mathit{assignmentWeights}[v:e] + \mathit{W}[v][e][f]$
                \State $\mathit{assignmentsPerFile}[f][v:e] = 1$
            \EndFor
        \EndFor
    \EndFor

    \State $\mathit{assignments} \gets \emptyset, \mathit{assignedVariables} \gets \emptyset$
    \State $\mathit{remainingFiles} \gets \mathit{F}$ \Comment{We ping-pong between shrinking this and $\mathit{compatibleFiles}$}

    \While{$|\mathit{assignmentWeights}| > 0$}
        \State $(\mathit{v^{OPT}}, \mathit{e^{OPT}}) \gets \argmax_{\mathit{a}} \mathit{assignmentWeights} [a]$
        \State $\mathit{assignments} \gets \mathit{assignments} + (\mathit{v^{OPT}}, \mathit{e^{OPT}})$

        \State $\mathit{compatibleFiles} \gets \emptyset$ \Comment{Recalculate which subset of $\mathit{remainingFiles}$ are compatible}
        \For{$\mathit{f} \in ($keys $\mathit{W}[v^{OPT}][e^{OPT}])$}
            \If{$\mathit{f} \in \mathit{remainingFiles}$}
                \State $\mathit{compatibleFiles} [f] \gets$ True
            \EndIf
        \EndFor
    
        \For{$\mathit{e} \in ($keys $\mathit{W}[v^{OPT}])$} \Comment{We've assigned $v^{OPT}$; remove it from $\mathit{assignmentWeights}$}
             \For{$\mathit{f} \in ($keys $(W [v^{OPT}][e])$}
                \If{$\mathit{compatibleFiles} [f]$}
                      \State \Call{Prune}{$\mathit{v^{OPT}}$, $\mathit{e}$, $\mathit{f}$, $\mathit{W}$, $\mathit{assignmentWeights}$}
                \EndIf
            \EndFor
        \EndFor
    
        \For{$\mathit{f} \in \mathit{remainingFiles}$} \Comment{Handle any $\mathit{remainingFiles}$ that are not compatible}
            \If{\textbf{not} $\mathit{compatibleFiles} [f]$}
                \State $\mathit{remainingFiles} \gets \mathit{remainingFiles} \setminus \mathit{f}$
                \For{$\mathit{(v,e)} \in ($keys $\mathit{assignmentPerFile} [f])$}
                    \If{\textbf{not} $\mathit{assignedVariables(v)}$} \Comment{Don't prune: already pruned early}
                        \State \Call{Prune}{$\mathit{v}$, $\mathit{e}$, $\mathit{f}$, $\mathit{W}$, $\mathit{assignmentWeights}$}
                    \EndIf
                \EndFor
            \EndIf
        \EndFor
        
        \State $\mathit{assignedVariables [v^{OPT}]} \gets True$
    \EndWhile
    
    \State \Return $(\mathit{compatibleFiles}, \mathit{assignments})$
\EndFunction
\end{algorithmic}
\end{algorithm}

\noindent\textbf{Complexity analysis of Algorithm \ref{algo:consistent_assignment_optimized}}. Assume each file has the same number of variables and candidate expressions.

\begin{enumerate}
\item Initial nested for loops to pre-calculate $\mathit{assignmentWeights}$ and $\mathit{assignmentsPerFile}$: O(vef)
\item Main while loop: We eliminate at least one variable per iteration, hence there are at most v iterations. Cost per iteration:
 \begin{itemize}
     \item Find the assignment ($v^{OPT}, e^{OPT}$) with the highest frequency: O(ve)
     \item Update $\mathit{compatibleFiles}$: O(f)
     \item Update $\mathit{assignmentWeights}$ to delete $v^{OPT}$: O(ef)
     \item Update $\mathit{assignmentWeights}$ to remove any incompatible files: \textbf{amortized} O(ve). Although this step is worst-case O(vef), each incompatible file is processed only once in total across the v loops.
 \end{itemize}
\end{enumerate}

The total cost of Algorithm \ref{algo:consistent_assignment_optimized} is:
  $O(vef)$ + $v * (O(ve) + O(f) + O(ef) + O(f) + O(ve))$ = $O((v^2)e + vef)$

\end{appendices}

\end{document}